# Future Wireless Network: MyNET Platform and End-to-End Network Slicing

Hang Zhang, Huawei Technologies Canada, Ottawa, Ontario, Canada,
E-mail: hang.zhang@huawei.com

## Abstract

Future wireless networks are facing new challenges. These new challenges require new solutions and strategies of the network deployment, management, and operation. Many driving factors are decisive in the re-definition and re-design of the future wireless network architecture. In the previously published paper "5G Wireless Network – MyNET and SONAC", MyNET and SONAC, a future network architecture, are described. This paper elaborates MyNET platform with more details. The design principles of MyNET architecture, the development of MyNET platform, the functions of MyNET platform, the automation of the creation and the management of end-to-end slices by MyNET, and the new capabilities enabled by MyNET are described in details.

# 1. Introduction

MyNET network architecture designed for future wireless network is described in this paper. In the previously published paper "5G Wireless Network – MyNET and SONAC" [1], MyNET and SONAC are described and discussed. This paper elaborates MyNET platform with more details.

It goes without saying, many driving factors are decisive in the re-definition and re-design of the future wireless network architecture.

## 1.1. Driving factors

- From the customer perspective

In 3G/4G, the dominate types of customers of a wireless network are subscribers who registered to an operator's network and used mobile devices to obtain the communication services. Such a subscriber could be an end-customer. In the future, more types of customers can be expected. In addition to subscribers, a wireless network operator needs to provide communication services to customers who associate with multiple devices distributed cross certain geographic area. Such customers include business customers who collect information remotely obtained from their devices, e.g., sensor, meters, robots, etc, and control these devices. OTT customers utilize wireless networks to provide services to their subscribers. Virtual network operators could be another type of customers of network operators. A virtual operator can provide communication services to its subscribers without owning a physical network and may involve in the operation of their virtual networks.

- From the service type perspective

In 3G/4G, the dominate services are voice, video, and message services. In the future, wide variety of types of services can be expected, such as, industry control services, remote surgery services, automatic care control services, drone control services, virtual life, etc.

- From the service requirement perspective

In 3G/4G, the limited QoS classes are defined for handling the traffic of subscribers. In the future, the huge disparate QoS requirements on rate, latency, and reliability could be expected.

- From the business model perspective

In 3G/4G, the players in wireless communication industry include the wireless network operators and end customers - subscribers. In the future, in addition to the subscribers, more players can be expected. Infrastructure network providers deploy infrastructure networks; wireless network operators control, operate and manage the infrastructure networks; virtual network operators could operate virtual networks which are provided by wireless network operators. The players and the working relationships among them form new business models in the future wireless network industry.

- From the terminal role perspective

In 3G/4G, a terminal is a device used by a subscriber for obtaining communication services. In the future, a terminal could be a part of infrastructure network on an on-demand basis. A terminal also could be a virtual entity which could include a group of sensors and move as a single entity.

- From the technical development perspective

New development of technologies and techniques in different technical fields are making inevitable impact on the design philosophy of the future wireless networks. Dig data, powerful computing capability, network function virtualization [2, 3, 4], SDN [5, 6, 7], etc will have profound and far-reaching impact on the control, management and operation of the future network.

- From infrastructure network resource perspective

In 3G/4G, the infrastructure network resources include communication equipment, such as, routers, switches, access nodes (cells), and connection links. The controllers are installed in secured center offices for the purposes of network control and management. In the future, the wireless network will be described as distributed cloud, including DCs and MECs (mobile edge computing) linked via high capacity networks. Such a wireless network could be denoted as a general wireless network infrastructure (GWNI) which not only includes traditional mobile network resources but also cloud resources.

Considering the aforementioned driving factors, the future wireless network architecture should:

- support a wide variety of service types - service customized slice [1, 8, 9]
- satisfy the huge disparate service requirements
- enable the automation of slice creation and adaptation
- enable the guaranteed slice performance during the lifetime of a slice
- enable simple and scalable slice operation after the deployment of a slice

- enable rapid GWNI adaption to GWNI performance degradation and traffic load change
- enable the openness to allow customers to define their slices, and even operate their slices
- enable the openness to allow 3$^{rd}$ party to involve in certain types of network management functions
- ensure the business revenues of all players
- make the architecture future-proof

By introducing the slice concept, a wireless network operator could become a slice provider who uses the integrated general infrastructure network from one or multiple infrastructure network providers to develop slices.

## 1.2. Design principles

In MyNET architecture design, three design principles are followed:

**Principle 1 – Separation of resource related functions (resource allocation/assignment) from other functions**

- Key functions – physical network resource allocation/assignment functions – should be exclusively controlled by slice providers
    - Customer or 3$^{rd}$ party can only manage their resource if the customer or 3$^{rd}$ party owns a hard slice (dedicated physical network resource)
    - Customer or 3$^{rd}$ party can only manage their resource at slice (virtual network) level
- Other network operation functions (NOS) are functions that do not directly involve in the allocation/assignment of resources but provide the required information to support the resource control and management
- Resource control and management includes two levels
    - Slice level: define resource allocation/assignment for slices and manage the multiplexing of the infrastructure network resource among multiple slices
    - End-points (devices/servers of a slice)level: manage the slice resource multiplexing among end-points of a slice

**Principle 2 – Modularization and re-organization of network NOS functions**

- Identify the required NOS functions by identifying the key types of objects which need to be controlled and managed
    - Device: reach-ability of devices, location and activity status of a device, need to be managed and the reach-ability information needs to be provided to the resource control and management entities.
    - Service: slices are 'network products' provided to customers by slice providers. After such a service 'product' is created, the operation and the performance assurance of the 'product' need to be managed.
    - Infrastructure network: infrastructure network is the physical resources from which all slices are created. The infrastructure network resources need to be monitored and managed to ensure the requirement performance. The resource availability information needs to be provided for further slice creation.

- Content: contents such as images, videos, web pages, etc, are contents which can be created by content producers and shared by remote interested parties though wireless network. Contents need to be managed to make the efficient content cache and sharing. This is similar to the management of publications in a library.
- Operational data: raw data related to the records of the traffic delivery status, equipment performance status, etc are the records of the network operation, control and management. These data needs to be managed to support the operations of other functions.

**Principle 3 – unified treatment of customer services and NOS services by resource control and management functions**

- Attributes of NOS
  - A specific type of services
  - Support of the optimization of the resource control and management of customer service slices by providing the required information
  - Sharable or dedicated (customized) to customers
  - On-demand exposure of information – internally and externally
- NOS services and customer services are treated in the same way by resource control and management functions
  - NOS slice can be jointly defined by both slice providers and customers via service description and requirement

These principles transform 4G control and management 'boxes' into flexibly recompose-able functions to best adapt to the new needs of 5G networks; enable efficient and customized network management and operation; transform 4G close network operation into controllable open operation environment.

## 1.3. Future network architecture - MyNET

Following these principles, MyNET architecture is design as shown in Figure 1.

- SONAC (Service Oriented virtual Network Creator)
  - SONAC-Com: responsible for the end-to-end slice composition, including software defined topology (SDT), software defined resource assignment (SDRA) and software defined protocol (SDP).
  - SONAC-Op: responsible for the end-to-end slice operation, i.e., for the delivery of device traffic over a deployed slice, including SDT-Op for the logical path management of device traffic, SDRA-Op for the physical resource assignment of device traffic and SDP-Op for the management of device traffic transmission protocol stack.
- Connectivity management (CM) – managed object: devices
  - Name based location tracking/resolution, activity status configuration and tracking; provisioning of the reach-ability information (where and when to deliver) of devices to SONAC.
- Customer service management (CSM) – managed object: customer services/slices
  - Slice (service )performance assurance

- o   Slice resource matching to the slice traffic
  - o   Slice operation assistance
- Infrastructure management (InfM) – managed objet: infrastructure networks
  - o   Infrastructure network performance assurance
  - o   Infrastructure network resource utilization optimization
  - o   Infrastructure network capacity abstraction
- Data analytics management (DAM) – managed object: operational data
  - o   On-demand and customized data log and analytics
- Content cache and forwarding management (CFM) – managed object: contents
  - o   Content registration
  - o   Content cache decision (what to cache and where to cache)
  - o   Content delivery (from where to deliver)

API interface is for the provisioning of the service description and requirement by customers/operators (slice providers).

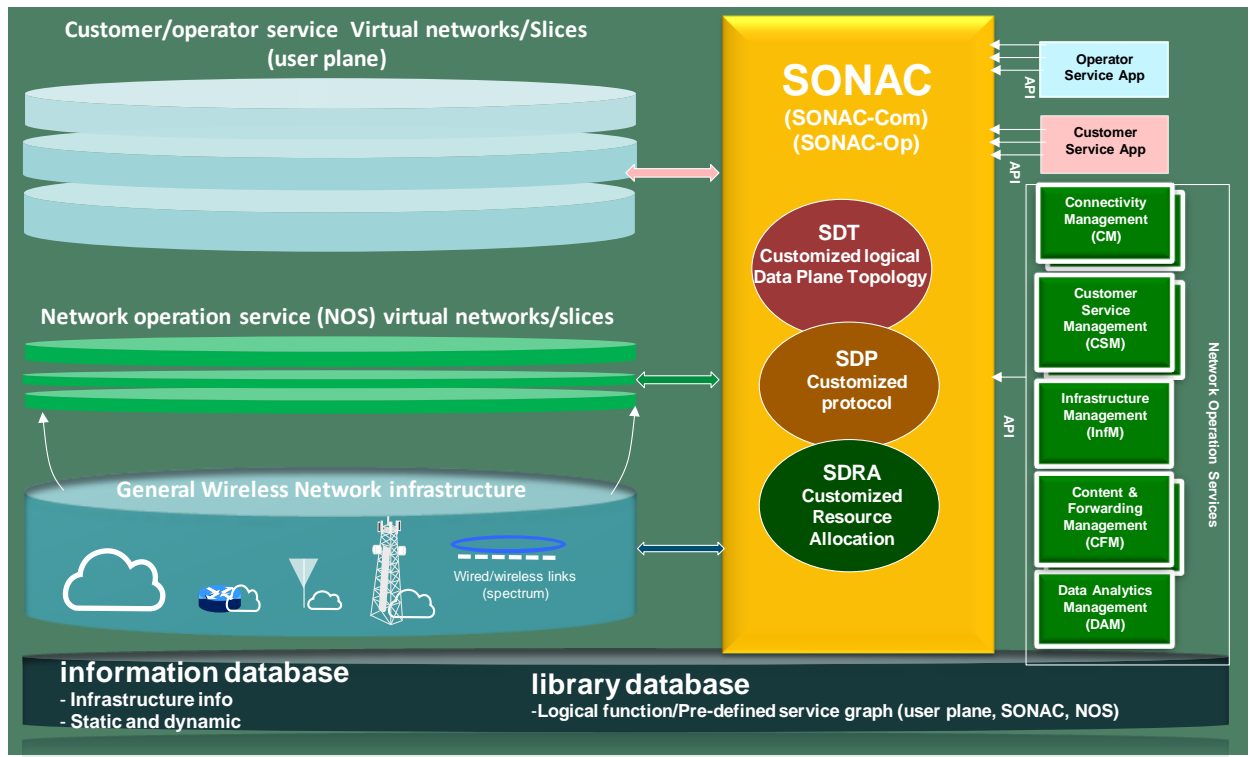

Figure 1. MyNET and SONAC.

## 1.4.   Terminology

For the easy description and discussion, the following terminologies are defined.

o   Slice providers who provides end-to-end slice using the integrated generic infrastructure resource

- Slice customers (consumers) who consume slice resource to obtain communication service (e.g., mobile subscribers)
- Slice customers (owners) who request and obtain from slice provider end-to-end slices and own the slices to provide further services to their customers (consumers), e.g., subscribers
- Infrastructure providers who provide infrastructure resource, e.g., DC, or transport network, or RAN, or spectrum, etc
- Intelligent functions that are programmed to enable the decision making based on inputs from API, or real-time conditions, etc
- End-points of a slice that are authorized to use a slice, including devices, servers, etc.
- User plane (UP) slice that consists of a set of resource for the delivery of data traffic of the end-points of a slice.

## 1.5. Example of general infrastructure network

In this paper, one example of the general infrastructure network is used, as shown in Figure 2. In this infrastructure network, there are a central (top) domain which includes a central DC and a network linking to two transport domains. The transport domains link the central domain to multiple RAN clusters (domains). A RAN cluster is consists of cloud (e.g., MEC), radio access network nodes with RF functions and full or partial baseband process functions. These access nodes are interconnected via wired or wireless backhaul links.

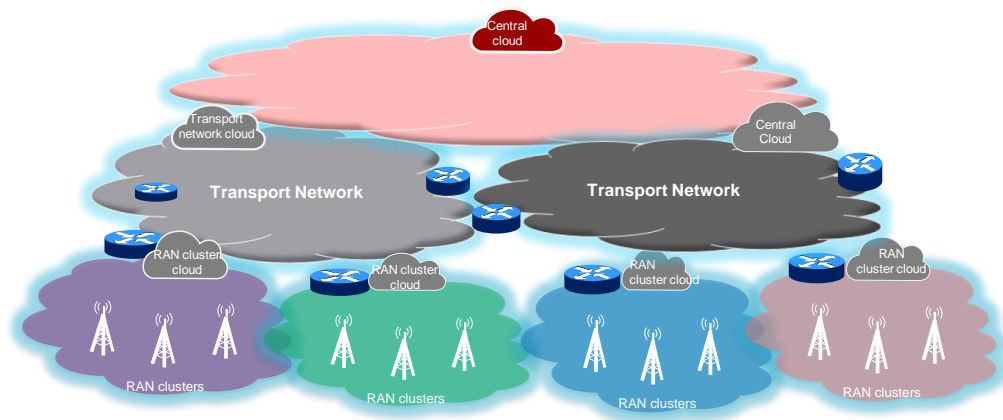

Figure 2. General infrastructure network.

## 1.6. Organization of this paper

This paper is organized as followings:

Chapter 2 provides the overview of the development of MyNET platform. Chapter 3 describes how to build up the libraries of MyNET platform. In chapter 4 and 5, the development and adaptation of MyNET platform are described. Chapter 6 describes the basic functions in the function libraries of MyNET platform. The development and adaptation of customer service slices are described in Chapters 7 and 8, respectively. Chapter 9 describes the operation of a slice. Chapter 10 describes how MyNET enables the business revenue assurance and Chapter 11 discusses how MyNET enables the openness. Conclusion of this paper is discussed in Chapter 12.

## 2. Overview of the development of MyNET platform

MyNET platform is developed to enable the automation of definition/development, deployment and operation of service customized slices using the integrated network infrastructure resources. The steps of the development of MyNET platform are highlighted in Figure 3.

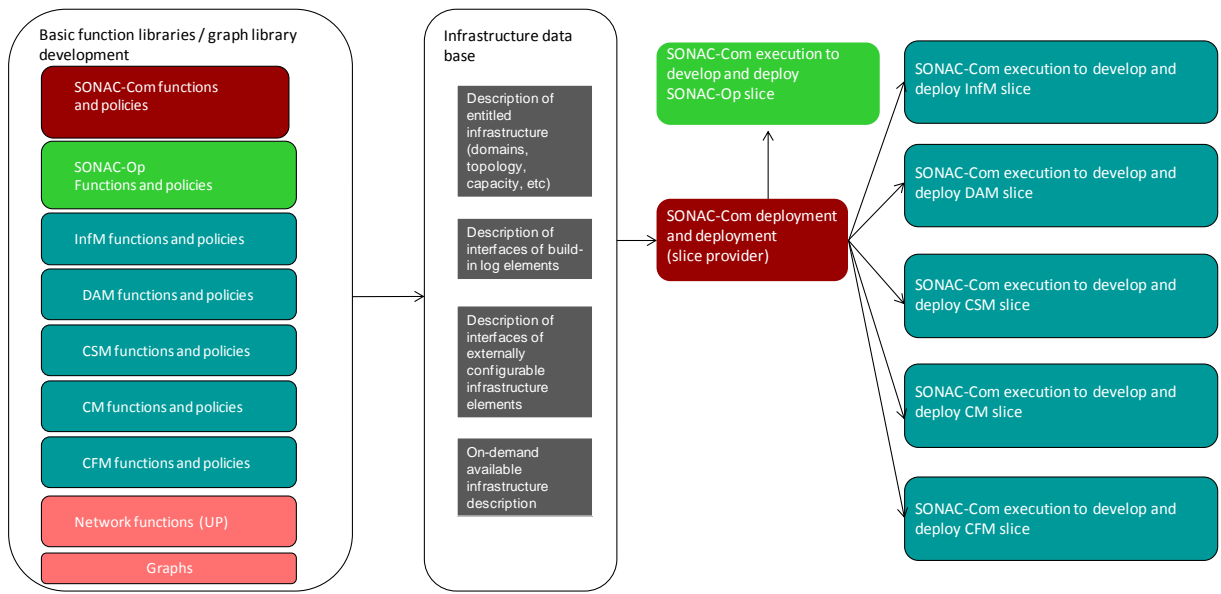

Figure 3. Development of MyNET platform.

To prepare MyNET platform, a slice provider needs to define (to program) the basic functions for MyNET platform. The slice provider also needs to provide the description of the topology, capacity and performance of the general infrastructure network. Then SONAC-Com functions are deployed into the infrastructure network by the slice provider. Following up the deployment of SONAC-Com, the execution of SONAC-Com functions will automatically develop SONAC-Op slice, NOS slices under the instruction of the slice provider via API interface as shown in Figure 1.

### 2.1 Development of basic function library and basic graph library

The development of basic function libraries is the most important step which defines the basic functions and the corresponding policies of intelligent functions. The functions defined and programmed include, SONAC-Com function family; SONAC-Op function family, InfM function family, DAM function family, CSM function family, CM function family, and CFM function family. The user plane network functions, such as mobile anchor point function, security functions, packet aggregation function, etc are also defined. In addition, the basic graph library is defined, which is used by SONAC-Com to determine the logical topology for slices, including NOS slices and customer service UP slices.

## 2.2. Infrastructure network database

To develop any type of slices using the infrastructure network resources, the information about the infrastructure network needs to be available to SONAC-Com.

- Description of entitled infrastructure network: provides information on the network hardware equipment/element and the physical connections among them. In MyNET platform, a DC or a MEC is abstracted as a single network node with certain process capability. For an infrastructure network integrated from multiple domains, the infrastructure description should provide the information of each domains and inter-domain connections.
- Description of interfaces of build-in log elements: provides information of interfaces which can be used to collect raw data from build-in log element developed by the infrastructure providers. These elements may be configurable by external entities to enable on-demand log. For example, a RAN node may have a build-in log element to log packet latency. This element may be configured to log latency of certain packets which only associate with an application, or a device or a customer or a slice, etc. Or a log element provides the log information with corresponding packet header info to a data analytics entity for information filtering and extraction, etc.
- Description of interfaces of externally configurable infrastructure equipments/elements: provides the information of interfaces which can be used by external entities to configure parameters for the operation of the equipment. For example, a RAN node can be configured to power-off certain sub-systems (e.g., access subsystem, wireless backhaul subsystem, etc), on an on-demand basis.
- Description of on-demand available infrastructure network: provides the information of on-demand infrastructure resource. A slice provider can establish collaboration with other infrastructure providers for the on-demand infrastructure resource integration. This is different from the entitled infrastructure resource. The entitled infrastructure resource is the infrastructure resource pool which is available all the time for the slice deployment even though the infrastructure resource pool could be the integrated resource pool from multiple infrastructure providers. The on-demand infrastructure resource is the resource which is integrated only if the entitled infrastructure resource pool cannot satisfy the requirements of new or existing slices, and at the same time, the infrastructure resource to be integrated is able to provide the required capacity. The corresponding information of interfaces as for the entitled infrastructure networks is also provided.

## 2.3. SONAC-Com deployment

SONAC-Com slice, i.e., the placements of SONAC-Com functions and the inter-connections among these functions over multiple domains, is determined by slice providers and deployed over the entitled infrastructure network. An integrated infrastructure network is usually composed of multiple technical domains, e.g., RAN domains, and transport domains, etc, which use different techniques for data traffic forwarding. A large scale infrastructure network is usually divided into geographical domains to make the control and management scalable, even though the entire infrastructure network belongs to a single owner. If the entire infrastructure network is an integration of multiple infrastructure networks the administrative domains become the natural domain division. The deployment of SONAC-Com is a decision of a slice provider given an entitled infrastructure network topology and capacity.

## 2.4. Development and deployment of SONAC-Op slice and NOS slices by SONAC-Com

SONAC-Op slice is developed and deployed automatically after SONAC-Com slice is deployed. The execution of SONAC-Com functions for SONAC-Op slice initialization develops SONAC-Op slice. For NOS slices, the NOS service description and requirement can be provided by the slice provider via API to SONAC-Com. The execution of SONAC-Com functions designed for developing NOS slices develops the NOS slices.

SONAC-Com is the key intelligent component of MyNET platform. After the definition/development, the design needs to be realized in the general wireless infrastructure network. The deployment of a well defined slice includes

- To instantiate, or activate or configure functions in the selected network nodes (e.g., data centers, mobile edge computing). This can be performed by ETSI VIM functions.
- To configure the inter-connections/interfaces among functions
- To configure infrastructure equipment for the mapping of slice logical tunnel onto physical infrastructure network
- To configure wireless network nodes (access nodes, wireless backhaul nodes) for the control and management of access link resource to enable an end-to-end slice
- Etc

Many standard bodies are actively defining these interfaces. E.g., ESTI NFV has done a lot of pioneer work in network virtualization and virtual functions instantiation, configuration and management. 3GPP is actively defining 5G network architecture [10].

## 2.5. MyNET platform

After the deployment of SONAC-Op slice and NOS slices, MyNET platform development is completed. Figure 4 shows the high level view of MyNET platform. The responsibility of each of function families is also marked.

- SONAC-Com receives requests from slice customers for slice development. SOMAC-Com interacts with the general wireless network to deploy customer service slices, which include UP slice, and may SONAC-Op and/or NOS slices if these slices need to be dedicated to customers. If there is no dedicated SONAC-Op and NOS slices are required, SONAC-Com configures the existing sharable SONAC-Op and NOS slices to enable the operation of customer UP slices.
- SONAC-Op, based on the configuration by SONAC-Com, controls the operation of slices for handling of traffic data forwarding over slices. It interacts with some of NOS functions, e.g., CM, CSM and CFM for the optimization of slice operation.
- InfM interacts with SONAC-Com to provide the information of the infrastructure network resource pool. InfM is also responsible for the infrastructure network monitoring, configuration to ensure the optimization of infrastructure network performance and resource utilization.

- CSM interacts with SONAC-Com for the slice performance assurance; interacts with SONAC-Op to support the optimization of customer slice operation; CSM also manages the slice charging to enable the business revenue assurance. In addition, CSM enables customized slice security.
- CM interacts with SONAC-Op to provide reach-ability information of devices for device traffic handling over slices.
- CFM is responsible for the optimization of Cache and Forwarding (CF) slice operation. It interacts with SONAC-Op for this purpose.
- DAM interacts with other NOS functions to provide data log and analytic services. DAM is a slice which directly interacts with infrastructure physical elements and virtual function elements for the purpose of data log.

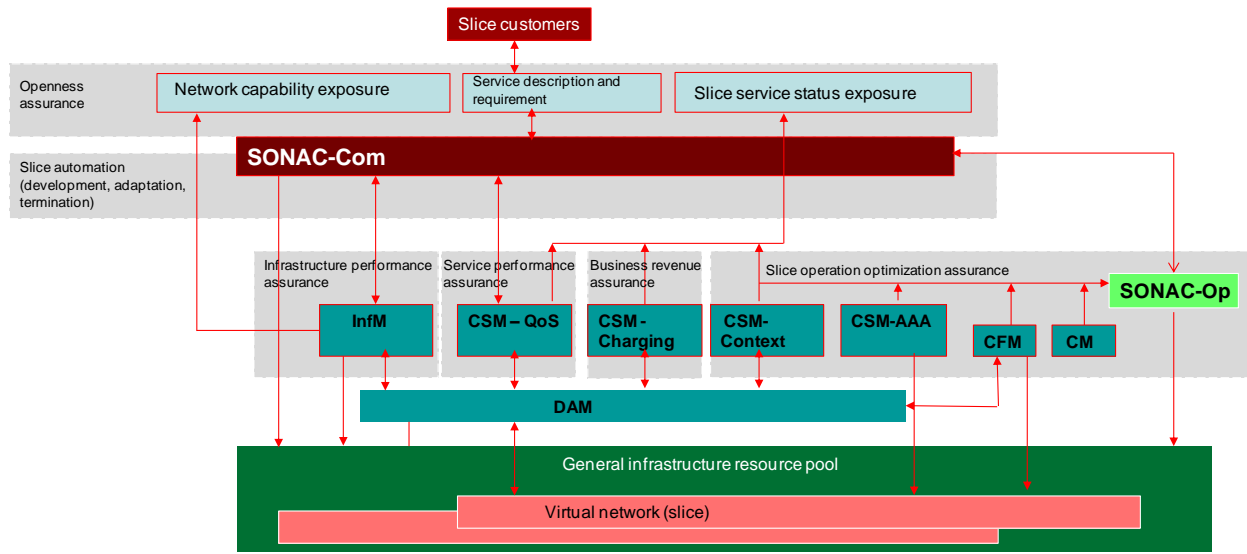

**Figure 4. MyNET platform.**

Figure 4 only shows the high level structure of MyNET platform and key interfaces. More details of each of these function families are discussed in Chapter 6.

## 3. Development of Function Library

The development of the basic function libraries of MyNET platform is to define (to program) functions and to make them to be selected and be deployed as needed.

### 3.1. Programming SONAC-Op functions and NOS functions

SONAC-Op functions are functions to control the traffic data delivery over deployed slices. For different types of services, different SONAC-Op functions can be defined. SONAC-Op performs different functions in different technical domains. To prepare the function library, SONAC-Op functions need to be defined. Among SONAC-Op functions, some of the functions need certain intelligence to make decision in order to react to certain real-time monitored conditions. For such intelligent functions, the rules or policies to govern the reaction of these functions need to be defined. SONAC-Op interacts with some of other NOS functions during the slice operation. The interfaces between SONAC-OP and other NOS functions need to be defined.

Similarly, NOS functions, for different technical domains and for different type of services, need to be defined. For intelligent NOS functions, the rules or policies to govern the reaction of these functions also need to be defined. To enable the required collaboration and interaction between NOS functions and SONAC-Op and among NOS functions, the interfaces should be defined. The definition of an interface processing function includes the definition on how to create a message for the interface and how to process a received message from the interface.

The basic SONAC-Op functions and NOS functions will form SONAC-Op function library and libraries of each of these NOS services. SONAC-Com will select appropriate functions from each of these libraries when SONC-Op slice and NOS slices are developed.

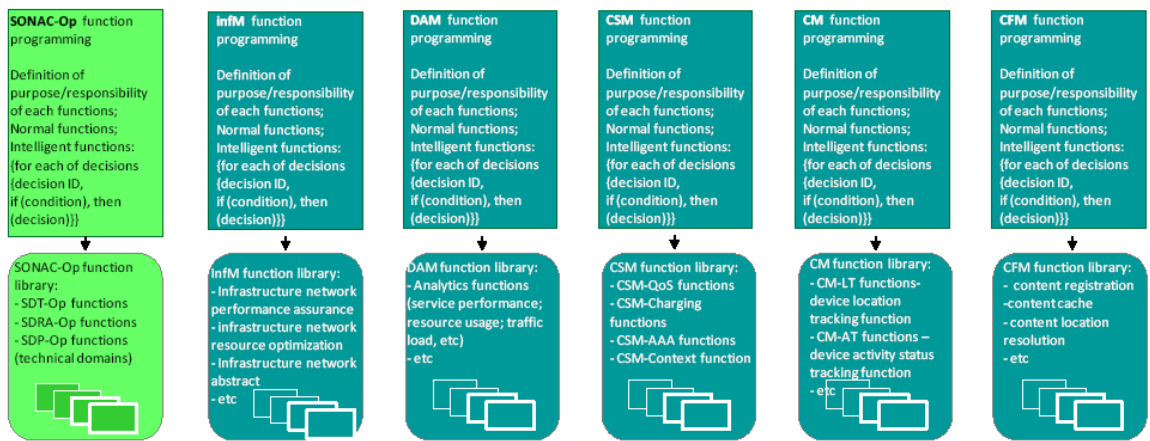

Figure 5. Programming SONAC-Op and NOS service functions.

Figure 5 shows the abstract expression of programming SONAC-Op and NOS functions.

In Figure 5, it is assumed that for each decision ID, an intelligent function needs to determine the decision based on 'conditions'. These rules, or called Policy, are pre-determined by slice providers.

## 3.2. Programming SONAC-Com

SONAC-Com is the key intelligent component in MyNET platform. It controls and manages the automation of slices, including SONAC-Op slice, NOS slices and customer service UP slices. The automation means the automatic slice development and deployment based on slice requests and description, and slice adaptation to real-time conditions without human involvement. From SONAC-Com perspective, both NOS services and customer services are viewed as the same. The only difference is that, for NOS slices, SONAC-Com selects the basic functions to build up a NOS slice from the corresponding NOS function libraries, while for customer service slices SONAC-Com selects the UP functions from the UP NF function library.

As shown in Figure 6, the main functions of SONAC-Com are categorized as follows:

- SONAC-Com functions for MyNET platform initialization and adaptation
- SONAC-Com functions for customer slice development
- SONAC-Com functions for customer slice automation/adaptation

In these figures, it is assumed that for an intelligent function, for each decision ID, the function needs to determine the decision based on 'conditions'. The 'decision' here is a generic term. SONAC-Com is the 'brain' of MyNET platform. In addition to take action (decision) based on the conditions, SONAC-Com also needs to have the capability of self-learning and real-time learning to adapt to variety of situations. These capabilities need to be programmed.

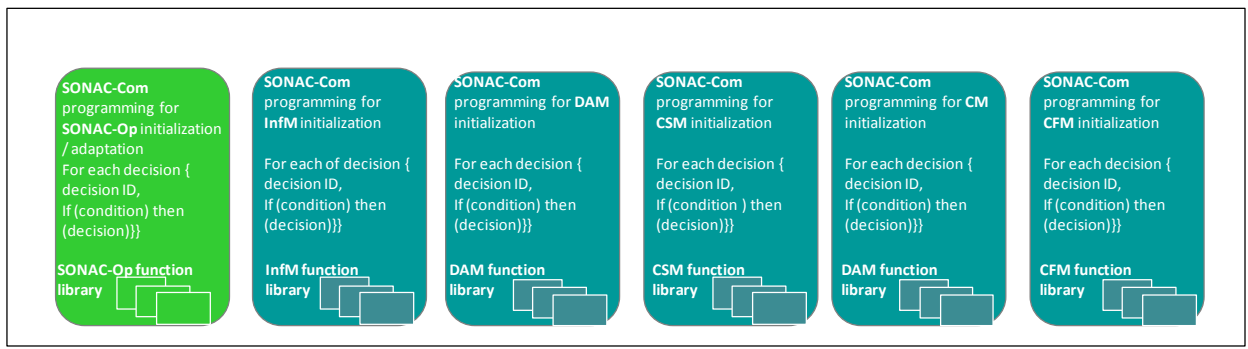

(a) Programming SONAC-Com for MyNET platform initialization.

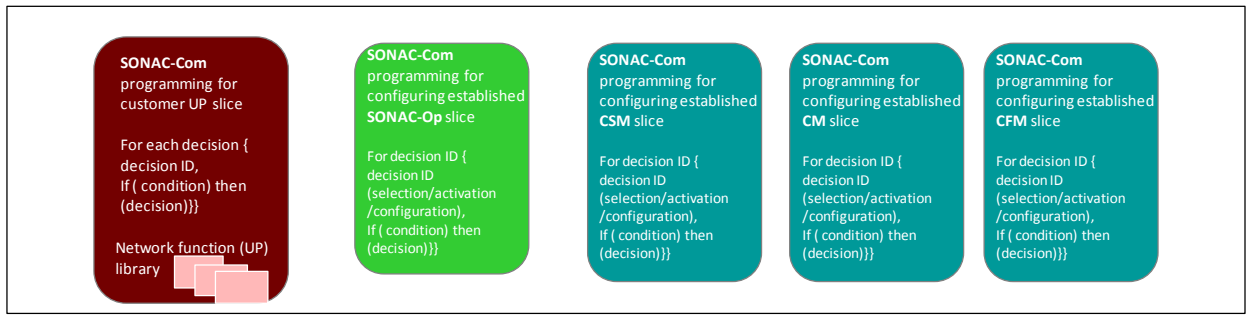

(b) Programming SONAC-Com for customer service UP slice development.

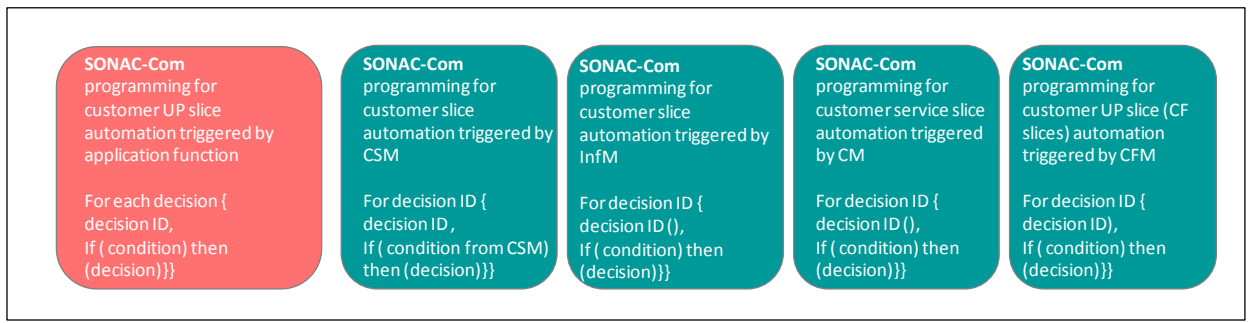

(c) Programming SONAC-Com for customer service UP slice automation/adaptation.

Figure 6. Programming SONAC-Com.

### 3.2.1. Programming SONAC-Com for MyNET platform development

For the development of MyNET platform, SONAC-Com is programmed for the initialization and adaptation of SONAC-Op slice and NOS slices, as shown in Figure 6 (a).

- Programming SONAC-Com for automation of SONAC-Op slice initialization and adaptation

SONAC-Com functions for SONAC-Op slice initialization and adaptation are defined and programmed such that SONAC-Com is able to develop SONAC-Op slice when MyNET platform is developed. SONAC-Com is programmed such that it is able to make the decisions on the selection of SONAC-Op functions from SONAC-Op library for different technical domains, on the determination of default parameter setting, on the definition of SONAC-Op slice topology, etc. The decisions also depend on the infrastructure topology and SONAC-Com slice topology. SONAC-Com also needs to be programmed to enable the adaptation of SONAC-Op slice to changes of infrastructure network topology.

- Programming SONAC-Com for NOS slices initialization and adaptation

Factors impacting the decision of SONAC-Com for NOS slices initialization and adaptation include the topology, technologies and capacity of the entitled infrastructure network. SONAC-Com functions are programmed for them to make decision on the selection of NOS basic functions from NOS function libraries for different infrastructure technical domains; to make the decision on logical topology of NOS slices; and to define interfaces to infrastructure network equipments/elements for InfM slice, DAM slice, etc. SONAC-Com is also programmed to enable NOS slice to adapt to changes of infrastructure topology. At the initialization of NOS slices, NOS functions are instantiated in the selected placements, some of the functions can be activated and some of them may be activated later at the development of customer service UP slices.

### 3.2.2. Programming SONAC-Com for the development of customer service UP slice

When a new slice request is received from a customer, SONAC-Com needs to develop a UP slice for the customer. For a customer UP slice, if existing sharable SONAC-OP and NOS slices can be used to support the operation of the UP slice, SOANC-Com needs to configure the involved functions in these SONAC-OP slice and NOS slices. If a customer requires a specific SONAC-Op slice or NOS slice (e.g., CM), SONAC-Com needs to determine the selection and activation of SONAC-Op or NOS functions implemented during the initialization of MyNET platform. SONAC-Com is programmed for this purpose, as shown in Figure 6 (b).

- Programming SONAC-Com for the development of customer service

Customer service slice (UP) needs to be automatically developed by SONAC-Com based on the topology and resource of the infrastructure network, and the service description and requirement. SONAC-Com functions need to determine the required network functions (NFs). In addition, the inter-connections among these NFs, and possible customer defined in-network application layer functions, need to be defined. The end-to-and protocol, slice logical tunnel protocol and access link protocols are also defined by SONAC-Com functions.

- Programming SONAC-Com for the configuration of SONAC-Op for a new customer service slice

At the development of a new customer service slice, SONAC-Com needs to select, and/or activate, and/or configure SONAC-Op functions which have been deployed at the initialization of MyNET platform. The decisions depend on customer service description, requirement and the customer service slice developed. SONAC-Com also needs to decide the required threshold(s) and parameters for functions if applicable.

- Programming SONAC-Com for the configuration of NOS for a new customer slice

At the development of a new customer slice, SONAC-Com needs to make decision on how to select and/or activate and/or configure some of NOS slices (e.g., CSM slice and CM slice). SONAC-Com needs to be programmed for this purpose. The factors which impact decision-making are the description and requirement of customer services and, in some cases, possible explicit requirement on NOS services by a slice customer.

### 3.2.3. Programming SONAC-Com for the adaptation of customer UP slice

SONAC-Com should be able to modify a customer UP slice after the slice is deployed. SONAC-Com needs to be programmed to enable this procedure, as shown in Figure 6 (c).

To enable the adaptation of customer service slices is an important aspect of MyNET platform and SONAC techniques. In MyNET platform, CSM-QoS functions, InfM functions, CFM functions, CM functions and application function within a customer service slice, may trigger the adaptation of an existing slice. SONAC-Com, thus, interfaces with these functions for this purpose. SONAC-Com needs to determine the slice adaptation based on the information provided by these functions. Decisions are made based on the received signaling messages, e.g., slice adaptation requests, from these functions. The reactions of SONAC-Com to these requests could be different depending on from which function an adaptation request is received.

### 3.3. Programming network functions for UP plane

In addition to the definition of SONAC-Com, SONAC-Op, and NOS functions, a slice provider also needs to define network functions for user plane. Such user plane functions could include mobile anchor, packet aggregator, packet dispenser, protocol translator, etc. Basic process units of transport protocol stacks in UP slices are also defined. Some of examples of the basic data process units are reliability assurance function, in-ordered delivery of packets, and security functions. When a customer service UP slice is developed, SONAC-Com needs to decide the selection and configuration of NF and protocols.

### 3.4. Pre-definition of basic graphs

The library of the basic graphs is defined by the slice provider. At the development of any types of slices, SONAC-Com needs to determine the selection of the required functions and to decide the end-to-end graph, i.e., the slice logical topology (interconnection among the functions) of the slice. There are multiple ways to define the graph of an end-to-end slice.

- For a relative small scale infrastructure network, an end-to-end graph may be pre-defined for each type of representative services. When the service request comes, SONAC-Com can decide the graph based on the service description and requirement.
- For a larger scale infrastructure network, the infrastructure network may be divided into multiple domains. Sub-graphs can be defined for each domain for each type of services. An end-to-end slice graph is the combination of all these sub-graphs.
- A slice provider can also only define basic graphs (a small number of graphs) for representative types of slices/services. The slice logical topology (end-to-end slice graph) will be obtained through running algorithms.

At the development of a slice, SONAC-Com will determine the graph, determine the mapping of the graph onto the general infrastructure networks which results the logical topology of the slice.

# 4. Execution of SONAC-Com for the initialization of MyNET platform

After SONAC-Com is deployed, SONAC-Op and NOS slices are developed and deployed by the execution of SONAC-Com functions.

## 4.1. MyNET platform initialization

SONAC-Com slice is developed and deployed by the slice provider. After this procedure, other slices can be automatically developed by SONAC-Com. Figure 7 shows the development of MyNET platform by the execution of SONAC-Com following the deployment of SONAC-Com.

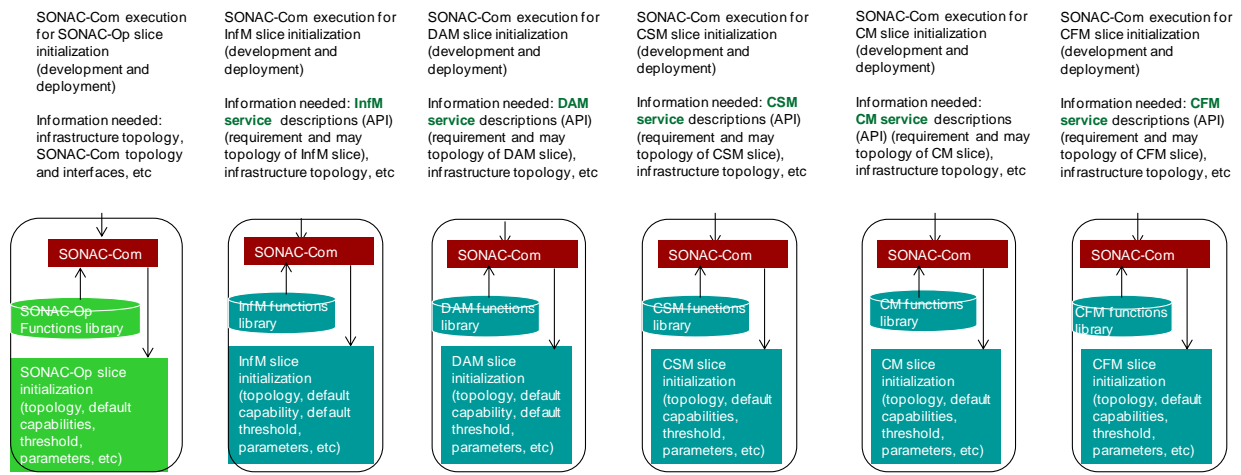

**Figure 7. Execution of SONAC-Com and MyNET platform development.**

- Execution of SONAC-Com function to develop SONAC-Op slice (initialization of SONAC-Op slice)

SONAC-Com should be able to develop and deploy SONAC-Op slice by the execution of the programmed SOMAC-Com functions. This procedure includes the decision making on SONAC-Op function selection for different technical domains, the placements of SONAC-Op functions, the interconnections among these functions, the configuration of the default thresholds/parameters of the functions, etc. the interconnection between SONAC-Com and SONAC-Op needs to be determined.

After the development of SONAC-Op slice, the slice needs to be deployed in the general infrastructure network. ESTI NFV defined VIM can be used for the instantiation of the functions in the selected locations (DCs, MECs, etc). SONAC-Com also needs to configure the interconnections of SONAC-Op functions. At the initialization, SONAC-Op functions may or may not be activated. SONAC-Op slice now is ready to be selected, activated and configured when a new customer slice is developed.

- Execution of SONAC-Com function to develop NOS slices (NOS slice initialization)

NOS service slices are developed by the execution of corresponding SONAC-Com functions based on the slice provider's instruction via API interface. At the development of a NOS slice, NOS functions for different domains are selected and the interconnections are defined. As for SONAC-Op slice, NOS functions are instantiated at the selected locations, and the inter-connections among functions of a NOS slice are established. The order of deployment and deployment of NOS slice can be as, InfM slice, DAM slice, CSM slice, CM slice and CFM slice. After the deployment of a NOS slice, SONAC-Op needs to be configured so that SONAC-Op can operate the slice. For each newly deployed NOS slice, the inter-connection between this NOS slice and previous deployed NOS slice needs to be established if needed.

## 4.2. Example of MyNET platform deployment

Figure 8 shows an example of a deployed MyNET platform for the purpose of illustration of MyNET platform deployment. In Figure 8 (a), a hierarchical architecture of SONAC-Com is shown. In this figure, the end-to-end SONAC-Com functions is deployed at the central cloud (e.g. DC), which inter-connects to SONAC-Com functions at two transport networks (transport network domains). SONAC-Com functions also are deployed in RAN clusters (RAN domains) in this example. A centralized SONAC-Com architecture is also possible where the end-to-end (global) SONAC-Com function is responsible for the end-to-end slice development over the entire infrastructure network.

Figure 8 (b) shows an example of deployed SONAC-Op and NOS slices.

SONAC-Op slice includes SONAC-Op functions deployed in domains' clouds, RAN clusters and access nodes, and is initialized by SONAC-Com.

NOS slices are also shown in Figure 8 (b) for the purpose of the illustration. SONAC-Com receives the description of a NOS service via API and then develops the NOS slice. In this figure, NOS slices functions are deployed in central cloud (DC), domain cloud and RAN cluster cloud.

We usually refer SONAC-Com as a SONAC-Com 'slice' since it includes a set of SONAC-Com functions deployed in multiple network locations (nodes) with established interconnections. Similarly SONAC-Op 'slice' and NOS 'slices' are also used in this paper.

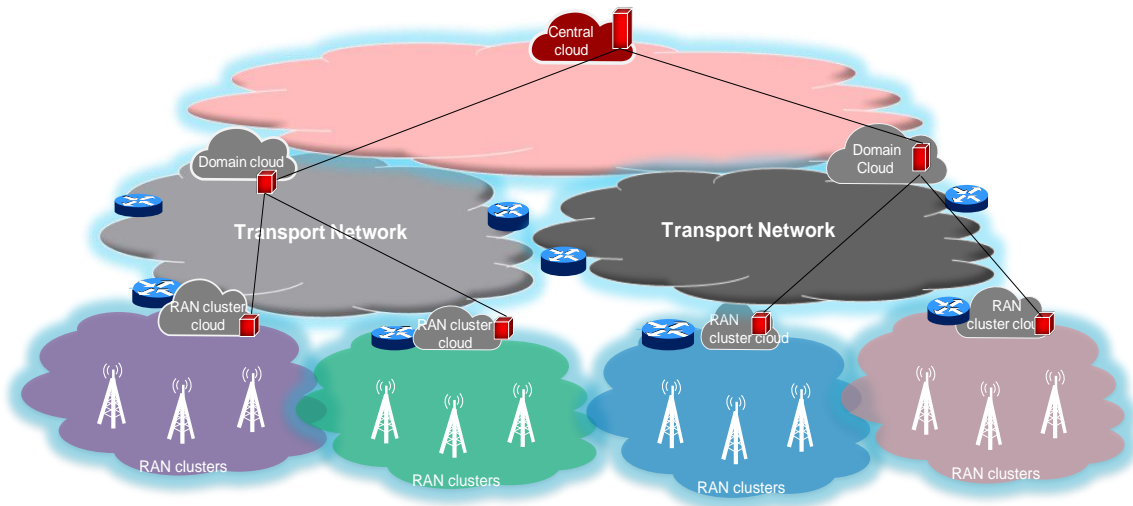

(a) Example of SONAC-Com Slice.

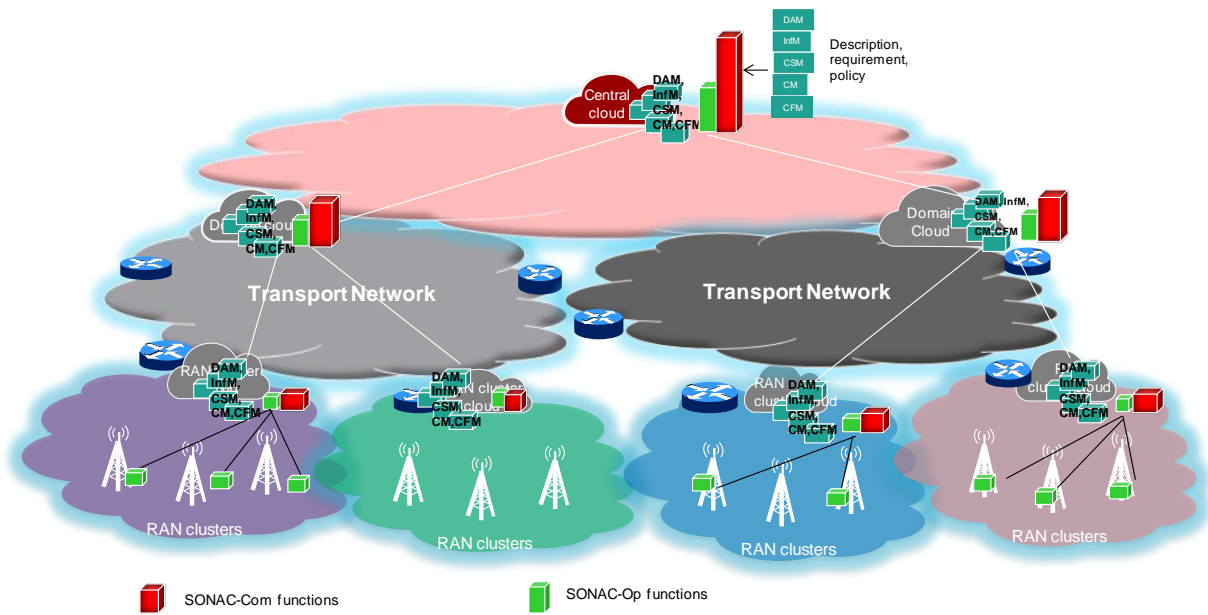

(b) Example of MyNET platform deployment.

**Figure 8. Example of MyNET platform deployment.**

## 5. Execution of SONAC-Com for the adaptation MyNET platform

MyNET platform should be able to adapt to the change in the infrastructure network.

This change could be caused by the integration of additional pieces of infrastructure networks into the entitled infrastructure networks when it is necessary. The degradation of the infrastructure network performance (equipment failure, etc) also causes the infrastructure topology change.

These changes could trigger the adaptation of MyNET platform topology. SONAC-Com needs to instantiate MyNET functions in newly integrated part or remove existing MyNET functions in some places. Although such event may not happen frequently, the automation of such adaptations by SONAC-Com is expected.

SOMAC-Com is programmed for this purpose, as described in Chapter 3.

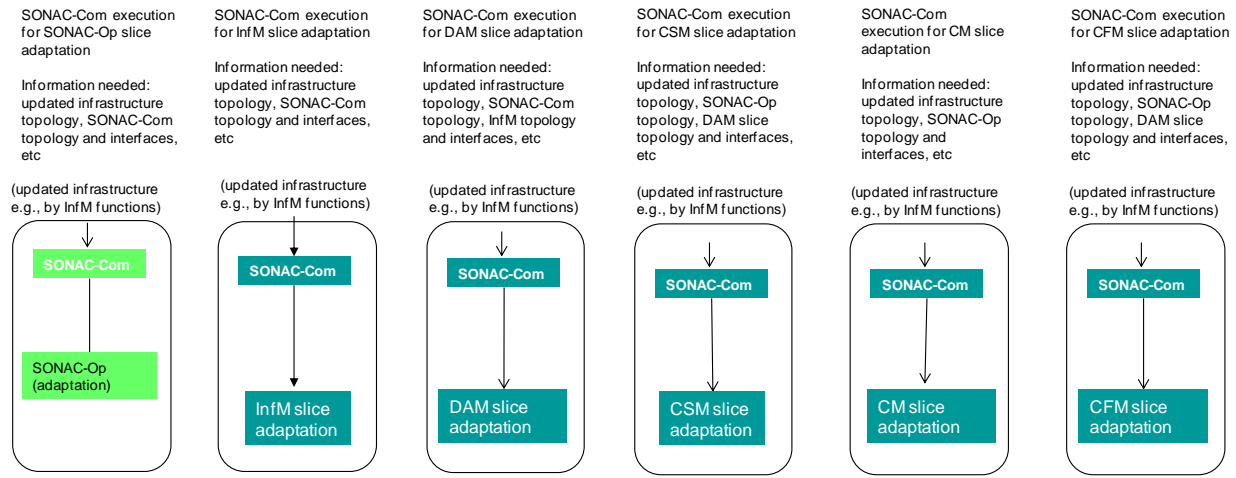

Figure 9. Execution of SOMAC-Com and MyNET platform adaptation.

Figure 9 highlights the execution of SONAC-Com functions for the purpose of MyNET platform adaptation.

Once a change in the current infrastructure network has been decided, the decision needs to be informed to SONAC-Com. Before the realization of a change in the infrastructure network, SONAC-Com needs to be aware of the change and make corresponding action, such as, to remove the current MyNET functions in the impacted domains. After the change has been realized in the infrastructure network, SONAC-Com is informed. SONAC-Com then and decides on the deployment of MyNET functions.

The quality of MyNET platform itself by a self-checking mechanism may also trigger the modification of the platform. This is not discussed in this paper.

# 6. Descriptions of MyNET platform functions

In this chapter, the key functions programmed during the development of MyNET platform are described. Please note that the functions described in this chapter are not a complete set of functions of MyNET platform. The purpose of this chapter is to provide a high level view of main responsibilities of MyNET platform components and relationships among them.

## 6.1. SONAC-Com functions

SONAC-Com is a newly introduced function family in the future network architecture. The key responsibilities of SONAC-Com include the creation, the adaptation and the termination of a slice. The key functions of SONAC-Com are the service coordination function, the resource coordination function and the resource management during the lifetime of a slice.

### 6.1.1. End-to-end SONAC-Com

In this segment, the key functions of the end-to-end service coordinator, the resource coordinator and the resource manager are discussed. Here a hierarchical architecture of SOMAC is assumed.

#### 6.1.1.1. End-to-end service coordination

The end-to-end service coordinator takes the responsibility of the provisioning of customer service slices over multiple domains.

***Functions:***
- Receive a request (QoS and service attributes) from a slice requester via API
- Analyze the request and determine which domains would be involved in the slice composition
- Analyze the request and provide the service requirement to each of domain service coordinators
  The end-to-end service requirement needs to be translated into the per domain service requirement. The determination of the per domain service requirement of a slice needs to made together with the resource coordinator and may need multiple iterations.
- Allocate a slice ID/service ID to a newly developed slice
  After a slice is developed and deployed, the set of resources (logical or physical) assigned to a slice is associated with a slice ID or service ID
- Receive performance reports from the end-to-end CSM-QoS
- Determine on the end-to-end slice resource modification

***Interfaces:***

- API for the requirement and attribute description of service slices (NOS services or customer services)
- End-to-end resource coordinator [service requirement, service description]
- Domain service coordinators [service requirement, topology graph, NFs]
- End-to-end CSM-QoS [slice performance]

#### 6.1.1.2. End-to-end resource coordination

The end-to-end resource coordination/management function is for coordinating the resource allocations among all domains. This function could be combined with the top (global) domain resource management.

There may be multiple schemes for the resource coordination. For a top-down scheme, top (global) domain (referring to Figure 8) may first determine the topology and resource allocation. Then the lower domains determine their resources for the requesting slice, given the design from top domain. However given such constrain, a lower (neighbor) domain may not be able to provide the required resource to the slice. In this case, the coordinator needs to coordinate the resource among domains. This may cause the upper domain to modify its previous decision. Similar strategy is applicable to other schemes, such as a bottom-up scheme.

*Functions:*

- Resource allocation coordination among domains

*Interfaces:*

- Domain resource managers
- End-to-end service coordinator

### 6.1.1.3. End-to-end resource manager or global resource manager

The end-to-end resource manager provides the slice resource allocation, based on the service description and requirement, and the available infrastructure network resource.

*Functions:*

- For slice definition/development
    - SDT-Com: determine the needed NFs (or NOS functions for NOS services), the end-to-end logical topology (slice level logical tunnels) across top domain or multiple domains or entire infrastructure network, depending on SONAC-Com architecture, the logical tunnel capacity requirement
    - SDRA-Com: determine the mapping scheme for the logical tunnels to the infrastructure network
        - for RAN cluster
        - for transport network
        - etc
    - SDP-Com: determine data transmission protocols
        - for end-to-end protocol
        - for RAN cluster protocol, access link protocol
        - for transport network protocol
        - etc
- For slice realization/deployment
    - SDT-Com: instruct the cloud resource managers for virtual function initialization, activation, configuration, etc
    - SDRA-Com: configure the forwarding rule for the involved infrastructure network nodes
    - SDP-Com: configure the protocol stacks for the involved infrastructure network nodes
    - Configure SONAC-Op and other NOS after a new customer service UP slice is developed
        - SDT-Com: configure SDT-Op: slice level routing table for routing traffic data over a slice; rule on end-point routing table establishment; rule on interaction with CM for location resolution of communication destination; the rule on end-to-end per device session establishment; slice logical topology (slice tunnel definitions)
        - SDRA-Com: configure SDRA-Op: rule on the resource assignment management for traffic forwarding over a slice, slice tunnel capacity, etc

- SDP-Com: configure SDP-Op: rule on protocol optimization for traffic flows under certain conditions
- etc

These functions are basic functions which are utilized for the development of customer service slice, SONAC-Op slice and NOS slices.

*Interfaces:*

- End-to-end service coordinator [service requirement], [domain ID, committed service delivery performance]
- End-to-end InfM [infrastructure resource MAP, available infrastructure resource MAP]
- Other domain resource coordinator/manager

Figure 10 shows the interfaces of end-to-end SONAC-Com. In this figure the end-to-end resource coordinator and global (top) domain resource manager are combined.

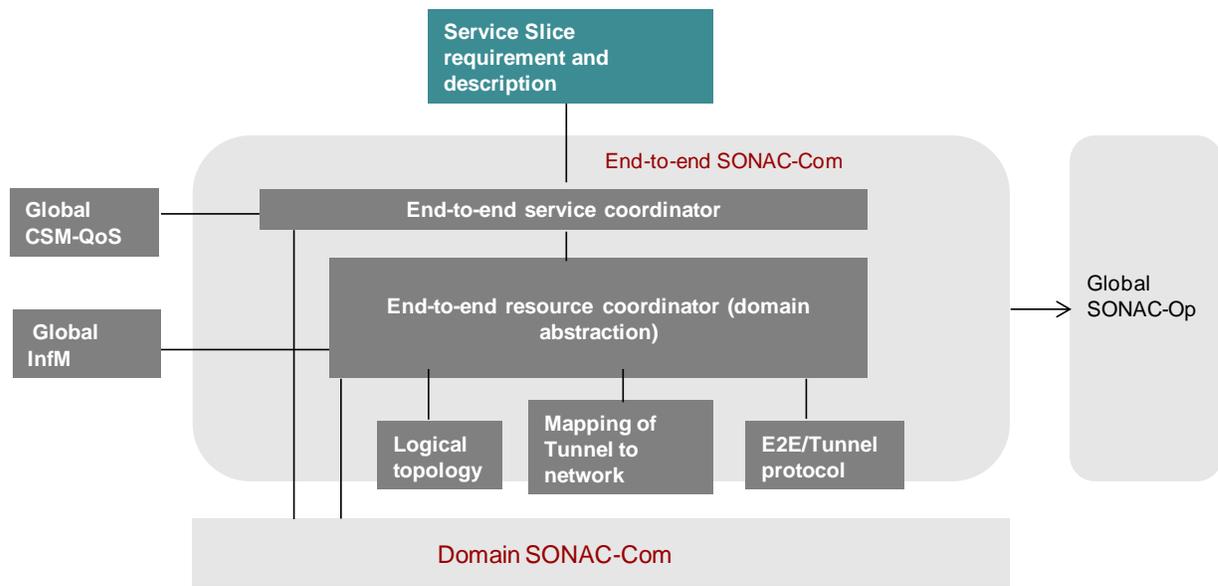

Figure 10. SONAC-Com functions: end-to-end service coordinator and resource coordinator.

### 6.1.2. Domain SONAC-Com

For an infrastructure network with multiple domains and a hierarchical SONAC-Com architecture is assumed, a domain may need a SMAC-Com function. A domain SONAC-Com will take the responsibility of the coordination and the resource management of the domain.

#### 6.1.2.1. Domain service coordinators

*Functions*

- Receive the service slice request from the end-to-end service coordinator or upper layer service coordinator
- Coordinate the intra-domain resource
- Receive the performance report from the domain CSM-QoS
- Determine on the slice resource modification in this domain

*Interfaces*

- end-to-end service coordinator or upper layer service coordinator
- domain resource coordinator
- domain CSM-QoS (slice performance assurance)

### 6.1. 2.2. Domain resource coordinator/manager

*Function:*

- receive the resource assignment decisions made by the end-to-end resource coordinator
- provide the domain infrastructure resource MAP to SDT-Com, SDRA-Com, and SDP-Com
- coordinate the domain resource (among SDT-Com, SDRA-Com, SDP-Com)
- slice design/development
    - SDT-Com: determines the logical topology of a slice on this domain
    - SDRA-Com: determine the mapping between logical tunnels to physical network on this domain
    - SDP-Com: determine the logical tunnel protocols, i.e., security, reliability, in-sequence transmission, etc on this domain
- Slice realization/deployment
    - SDT-Com: instruct the cloud resource managers for virtual function initialization, activation, configuration, etc
    - SDRA-Com: configure the forwarding rule for the involved infrastructure network nodes
    - SDP-Com: configure the protocol stacks for the involved infrastructure network nodes
    - Configure SONAC-Op
        - SDT-Op
            - Slice logical topology (slice tunnel definitions)
            - Slice level routing table
            - Rule on end-point routing table establishment
            - Rule on interaction with CM for location resolution of communication destination
        - SDRA-Op
            - Rule on packets delivery over logical tunnels of a slice
            - Rule on access link scheduler
            - Etc
        - SDP-Op

- Rule on protocol adaptation of traffic packet delivery under certain condition within the scope of per slice protocol set
- etc

The domain SONAC-Com functions depends on the domain technology. For example, for a RAN cluster, the domain SONAC-Com takes the responsibility of wireless backhaul and access link resource allocation and configuration of the access nodes, backhaul nodes, etc.

*Interfaces:*

- End-to-end resource coordinator [determined and recommended NFs, graphs, etc, by the end-to-end resource coordinator]
- SONAC-Op [configurations]
- Domain InfM [infrastructure resource MAP, available infrastructure resource MAP]
- Other domain resource coordinator

The interfaces of a domain SONAC-Com is shown in Figure 11.

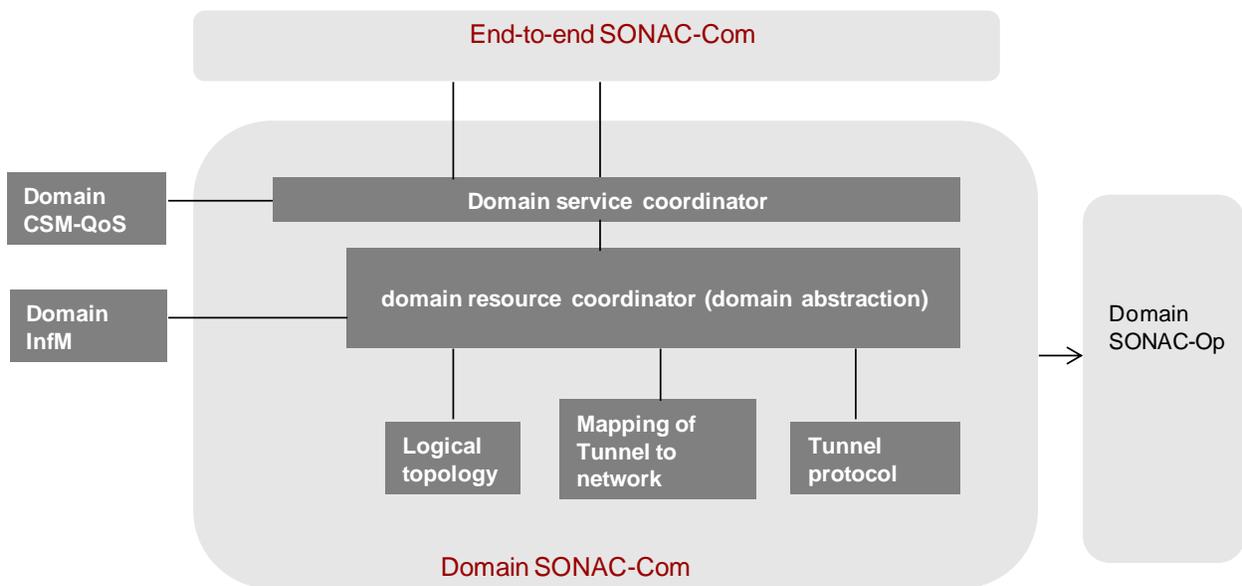

Figure 11. Doman SONAC-Com functions: domain service coordinator and resource coordinator.

## 6.2. SONAC-Op functions

SONAC-Com determines the multiplexing of the infrastructure resource among slices, i.e., SONAC-Com develops slices using the general infrastructure resource pools, while SONAC-Op is managing the multiplexing of a slice resource among devices for the delivery of the traffic packets over the deployed slices. SONAC-Op communicates with CSM-QoS functions to ensure the per device/application performance. SONAC-Op communicates with CSM-Context functions for the optimization of the device traffic delivery. SONAC-Op interacts with CM functions for the purpose of the routing and delivery of device packets. SONAC-Op functions need to be designed for different technical domains, e.g., RAN domains and transport domains, etc. SONAC-Op functions defined depend on the format of the slice resource assigned. The formats of the slice resource assignments are discussed in Chapter 7. Here the key functions of SONAC-Op are discussed.

***Functions:***

- SDT-Op
    - Determine the path for traffic packets at slice (virtual network) level
        - Work as virtual routers if a slice is defined with a completely slice logical topology and tunnel capacity definition
        - Manage the end-to-end per device session establishment if a slice is designed without a complete slice topology definition
    - Receive the slice level routing table from SONAC-Com
    - Create and maintain the end-point routing tables
    - Interact with CM for the purpose of packet routing over a slice or per device session establishment
- SDRA-Op
    - Determine the physical path given a logical tunnel for traffic packets
    - Access link resource access control cross slices with no dedicated access link resources assigned (RAN domain)
    - Access link resource access control for slices with the dedicated access link resource allocated (RAN domain)
- SDP-Op
    - Determine the protocols given a logical tunnel for traffic packets if the default slice protocol is not optimal
    - Access link protocol (RAN domain)

Some of slice requires service specific SONAC-Op. For example, SONAC-Op functions supporting a Content cache and Forwarding (CF) slice includes:

- SAT-Op
    - Content request routing
    - Response routing
    - Maintenance of response routing table

***Interfaces:***

- SONAC-Com [configurations]
- CM to maintain end-point routing table [device ID, current anchor, etc]
- CSM-QoS [device ID, performance, etc]
- CSM-Context [device ID, behavior ID, etc]
- CFM [content ID, cache ID]

Figure 12 shows SONAC-Op functions and the interfaces.

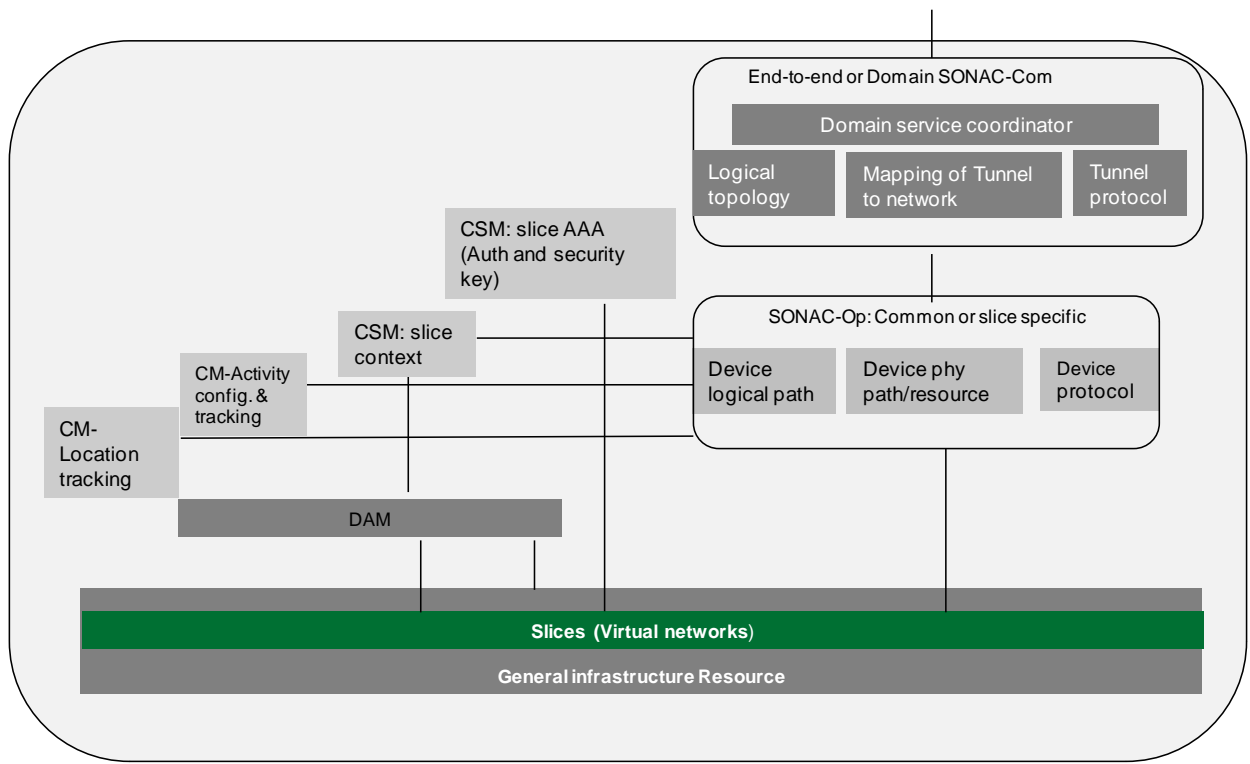

Figure 12. SONAC-Op functions and interfaces.

## 6.3. InfM functions

In 3G/4G, the entire physical infrastructure resource can be viewed as one slice which provides MBB service to subscribers. The management of such an infrastructure consists of configuration, failure, performance, etc.

In the future network, the general infrastructure network resource includes not only mobile network resource but also the cloud resources distributed cross the entire networks, as described in Chapter 1. Due to the introduction of the network slicing concept, the general infrastructure network resource needs to be multiplexed by multiple service customized slices. The management of such an infrastructure network is facing new challenges. It is expected that the future infrastructure network resource could be 'elastic' to adapt to the changes of the infrastructure network performance and the traffic loads. In order to enable the network slicing, the available resources of the infrastructure network needs to be provided to SONAC-Com for the purpose of the slice development. In some cases, the abstraction of certain network domain is needed to simplify the operation of SONAC-Com. For example a DC is abstracted as one node with certain processing capability; one RAN cluster can be abstracted as one node with certain traffic delivery capability and with one or more ports connecting to other domains.

Figure 13 shows the interface of infM functions.

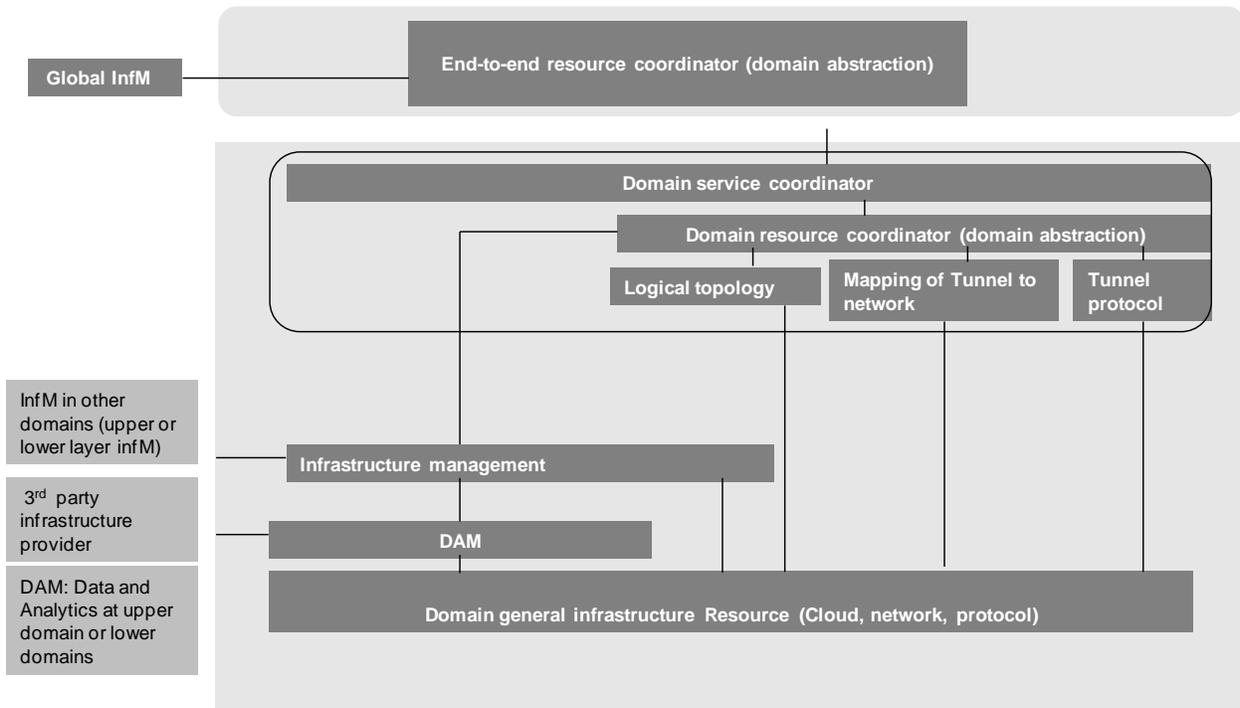

**Figure 13. InfM functions and interfaces.**

The key responsibilities of InfM are listed as follows.

*Functions:*

- Configure DAM for the periodic performance monitoring of the infrastructure network (DC, MECs, network nodes, access node, etc) for the infrastructure performance assurance
- Configure DAM for periodic monitoring of the available/remaining resource (DC, MECs, link, access link, and spectrum)
- Configure DAM for the periodic monitoring of the infrastructure resource utilization
- Determine on the infrastructure topology modification (e.g., wireless BH configuration)
- Determine on the infrastructure configuration (e.g., turn on and turn off the infrastructure elements)
- Determine on the over-the-air interface (e.g., frame structure)
- Determine on the 3rd party infrastructure integration (3rd parties on-demand available infrastructure topology is pre-stored)
- Negotiate with 3rd party infrastructure providers
- Manage the integration of device as one network node
- Perform domain resource abstract to upper layer
- etc

*Interface:*

- DAM for configuring log and analytics [location ID (where to log), information ID, reporting mode, log mode]
- Domain resource coordinator [domain infrastructure MAP, available resource MAP]
- InfM in other domains to provide the domain resource abstract [domain abstract MAP]
- 3rd party infrastructure resource providers for the infrastructure integration [requesting infrastructure description MAP]
- Infrastructure elements (access nodes, etc) for the element configuration[configuration parameters]

InfM functions need to be designed for different technical domains. For example, the access network node configuration function is only applied to RAN cluster. The abstraction functions of InfM for DC cloud and transport network are different.

## 6.4. DAM functions

In 3G/4G, certain level of data log and analytics are used for the purpose of managing the infrastructure network and for the purpose of charging. In the future network, it is expected to utilize the techniques, such as Big Data to better manage and operate the network. It is also expected that a unified data analytics service to provide the service for a variety of the operation and management purposes. DAM function family is introduced for this purpose.

Figure 14 shows DAM function and the interfaces.

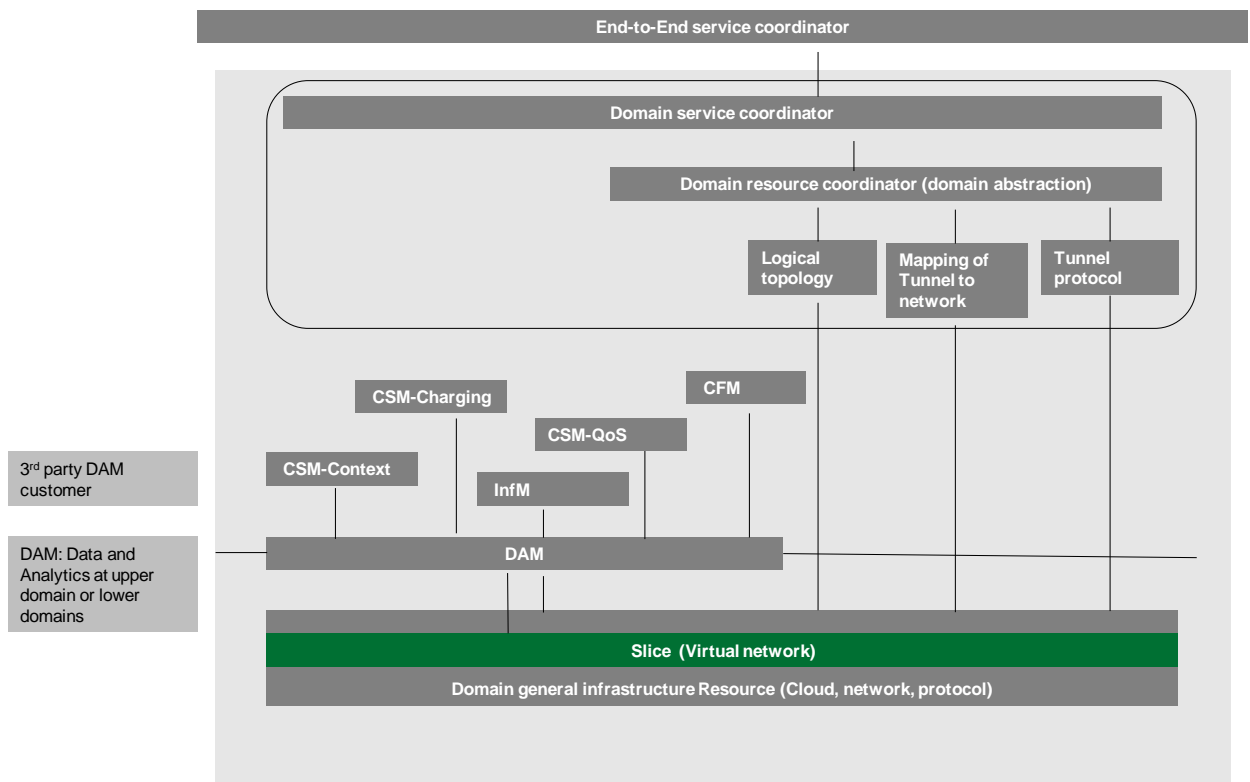

Figure 14. DAM functions and interfaces.

The key responsibility of DAM is the on-demand data log and analytics. DAM slice is the slice which utilizes the build-in log elements in infrastructure equipment elements and the virtual log elements created for logging virtual network functions for the purpose of the raw data log and analytics. DAM slice is a unified log and analytics slice to provide services to other NOS services and 3$^{rd}$ parties.

To obtain DAM service, the requester needs to indicate the follows:

- Log mode:
    - periodical monitoring and log (configurable period)
    - on-demand monitoring and log
    - etc
- Report mode:
    - threshold based report
    - periodical report
    - poll based report
- Log ID:
    - packet process time
    - number of packets over a window
    - number of bytes over a window
    - etc
- Log attributes ID:
    - application ID (all packets associated to an application)
    - session ID (all packets associated to a session of a device)
    - subscriber ID (all packet associated to a subscriber/device)
    - service customer ID (all packets associated to customer)
    - slice ID (all packet associated to a slice)
    - etc
- Information ID:
    - average packet latency
    - throughput
    - etc
- etc

*Functions:*

- log and analyze data
- provide the required information

*Interfaces:*

- CSM-QoS [slice resource utilization statistics; slice traffic load distribution statistics; service performance statistics]
- CSM-Charging [subscriber ID and charging info; slice ID and charging info; slice provider ID and charging info]

- CSM-Context [slice ID, context ID, context info]
- InfM [infrastructure performance MAP, available/remaining resource MAP, resource utilization MAP]
- CFM [interesting rate distribution of contents]
- DAM in other domains
- 3[rd] party DAM customers

Different domain techniques may require different DAM techniques. When a NOS service needs certain information, the NOS needs to indicate the corresponding NOS function ID such that DAM can provide the required information.

## 6.5. CSM functions

CSM function family is introduced dedicated to the slice related control and management. A slice can be viewed as a 'network product'. After the product is delivered to a customer, an entity should take care of the performance of the 'product' and make the 'product' satisfy the requirement, etc. CSM function family is designed for the purpose. The responsibility of CSM functions is to manage a deployed customer service slice. The functions of CSM are categorized into a number of sub CSM function families: CSM-QoS, CSM-charging, CSM-AAA, CSM-context.

### 6.5.1. CSM-QoS

The responsibility of CSM-QoS is to ensure the performance of a slice, the performance of a customer of a slice and the performance of per subscriber in case applicable. CSM-QoS utilizes a close-loop mechanism to ensure the 'elastics' of the resource of a customer service slice to adapt to the change and dynamics of the slice traffic. After a customer slice is deployed, the pre-deployed CSM-QoS (during the initialization of CSM slice) needs to be configured, including the selection, activation and configuration of CSM-QoS functions. Or a dedicated CSM-QoS function/slice can be instantiated if needed. CSM-QoS can trigger the slice resource adaptation based on real-time monitored status of the slice resource utilization, the slice performance, and the slice traffic load pattern. CSM-QoS interacts with DAM for obtaining the status information.

*Functions at slice level*:

- Configure DAM to monitor the slice resource utilization (for the best matching of the slice resource and the slice traffic)
- Configure DAM to monitor the slice traffic load change and migration
- Configure DAM to monitor the slice QoS (for slice service performance assurance)
- Configure DAM to monitor slice traffic change pattern - time dependency and geographic dependency (for better planed resource allocation to a slice over time and geographic areas)
- Determine on the slice resource modification based on the monitored results
- Report the performance on an on-demand basis

*Functions at service level (assuming that a slice supports multiple service customers and a customer associates with multiple devices)*

- Configure DAM to monitor the per performance of a service customer
- Configure DAM to monitor per service (customer) traffic change pattern
- Report per service level performance and service traffic load statistics

*Functions at subscriber level:*

- Configure DAM to monitor the per subscriber traffic performance
- Report the per subscriber traffic performance

*Interfaces:*

- DAM for configuring log and analytics [location ID (where), information ID (what), reporting mode ID; log mode ID]
- Service coordinator for the slice performance assurance [slice ID, delivered slice performance]
- Resource coordinator [slice ID, slice resource utilization ratio]
- CSM-QoS in other domains and end-to-end service coordinator and end-to-end resource coordinator) [domain ID, slice performance, slice resource utilization, slice traffic load statistics]
- Service coordinator for close loop slice traffic characteristic estimation [traffic load dependency on time, traffic load dependency on geographic location]

Please note that the information content carried in these interface are not a complete set of contents.

### 6.5.2. CSM-Charging

In 3G/4G, the charging function is basically for the billing purpose. The log function is located in PGWs and the charging rule is determined by the operator. Due to more players involved in the future network, the design of the charging functions needs to make sure to include all the players for the business revenue assurance. CSM-Charging functions are designed for this purpose. The charging function will provide changing information in multiple granularities, at the level of per application of a device/subscriber, at the level of a device/subscriber, at the level of slice customer (with multiple devices or subscribers), at the level of a slice provider. CSM-Charging function should enable

- Multiple level granularly charging
- Service customized charging function
- Configurable location of the log function for the purpose of charging
- Both BW and Cloud consumption based
- Both the delivered traffic based and the reserved resource based
- Negotiable billing policy
- Charging rule update based on network load/resource
- SLA model and Per-pay-per-service model
- Configurable placement of charging-related function
- Both off-line and on-line mode

CSM-charging functions utilize DAM function to obtain the information for various charging purposes. CSM-Charging function is configured by SONAC-Com at a new customer service is deployed.

*Functions (application level, subscriber level, service level, slice level, etc)*:

- Configure DAM for logging the charging related data (where to log and what to log), and what information need to be extracted from the logged raw data)
- Report the charging information on demand

*Interfaces:*

- DAM for configuring log and analytics [location ID (where to log), information ID, reporting mode, log mode]
- CSM-charging functions in other domains

### 6.5.3. CSM-AAA

In 3G/4G, a subscriber needs to register to its operator before using the network of the operator. After the authentication/authorization, the subscriber is authorized to use the network. The traffic of the subscriber is securely protected in the over-the-air interface.

In the future network, due to the introduction of the slice concept, a device may need per slice authorization and each of a slice may have the service customized authorization schemes, the service customized key management schemes, etc. After a slice is created, a control entity is needed to control the authorization of devices to use the slice, and manage the keying materials of this slice. CSM-AAA is designed for this purpose. The responsibility of CSM-AAA function is to enable the efficient and customized device AAA and key management. After a customer slice is developed, SONAC-Com needs to configure the sharable CSM-AAA function deployed during the initialization of CSM slice, or defines a dedicated CSM-AAA function/slice for the newly deployed slice.

*Functions:*

- Manages the slice specific authentication and authorization of devices for devices to use a slice
- Determine the key material distribution to the security functions of the slice
- Maintenance of profiles of devices if needed

*Interfaces:*

- Slice security function for configuring the security key of device/slice [device ID, security materials]
- CSM-AAA functions of other domains [key management (crosses domain)]

### 6.5.4. CSM-Context

The responsibility of CSM-Context function is to utilize DAM to log per slice-wise service characteristics or per device/subscriber behavior to assist a customer for the better understanding of the service attributes of its devices/subscribers. This can be used to assist SONAC–Op for the better management of device/subscriber traffic forwarding.

*Functions at slice/service level:*

- Configure DAM for the slice service characteristics statistics (distribution of application/service requested, etc)

*Functions at device level:*

- Configure DAM for monitoring the device behavior (e.g., commute pattern, application/service preference, frequent communication parties/partners, etc)

*Interfaces:*

- DAM for configuring log and analytics [location ID (where to log), information ID, reporting mode, log mode]
- SONAC-Op for slice resource utilization optimization [device ID, context ID and data];
- Slice customer for slice service characteristics

Figure 15 shows CSM functions and the interfaces.

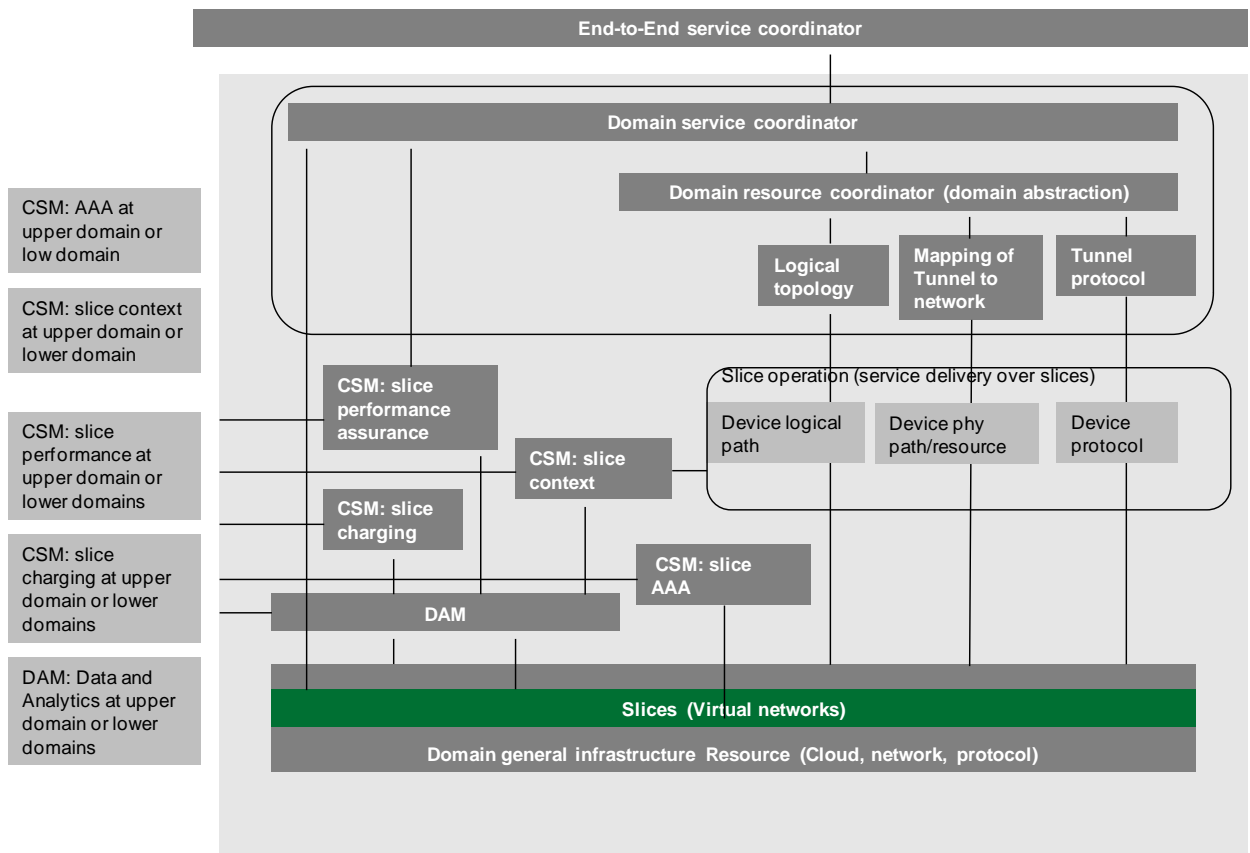

**Figure 15. CSM functions and interfaces.**

## 6.6. CM functions

In 3G/4G, a subscriber is tracked through handover procedure and location updates by the centralized MME (mobility management entity). Since the subscribers are mobile users, the location is unpredictable and the downlink traffic arrival is also difficult to be predicted (e.g., incoming call), the network needs to perform the location tracking, configure the activity state of devices and maintain the information in order to deliver the downlink traffic to a device in the right 'place' and at the right 'time'.

In the future network, a hierarchical architecture of the location tracking/resolution management is preferred to enable the local routing. Decoupling name with location of a device is expected to reduce the complexity of the traffic delivery caused by the location change of a device. Due to the wide variety of service types, the location tracking/resolution and the activity state control and management can be different for different type of services. CM function family is designed to enable the above.

The key responsibility of CM functions is to provide the reach-ability information of devices to SONAC-Op for the purpose of device traffic delivery over slices. The important principle of CM is to decouple the name and its location of a device. The architecture of CM slice presents a hierarchical structure which makes the reach-ability management scalable, in addition to enable the local routing. In all layers of a CM hierarchy, the location information of a device is associated with the name of the device, and its current "location" could be the ID of an access node, or the ID of a RAN cluster or the ID of a virtual anchor point function. Two types of functions are considered. The device location tracking (LT) function is to track device location and provide the information to SONAC-Op for the packet routing purpose (where to deliver packet to a device). The device activity configuration and tracking (AT) function is to configure and to track the activity status of a device. The AT function provides the information to SONAC-Op, e.g., scheduler at access node for it to schedule DL traffic delivery (when to deliver packets to a device). In the future network, due to the variety of types of services, CM needs to be designed to be customized to the attributes of services and devices.

*Functions of CM-LT:*

- Device location tracking and resolution (common or customized)
    - UL based LT
    - DL measurement based LT
    - LT with predictable DL packet delivery
    - LT for UL transmission dependent DL packet delivery
    - LT with customer assistance
    - etc

*Functions of CM-AT*:

- Device activity configuration/tracking
    - Customer assistant activity status configuration
    - Customer assistant activity status tracking
    - etc

*Interfaces:*

- SONAC-Op (SDT-Op) [device ID, location (domain ID, or cluster ID or access node ID)];

- SONAC-Op (SDRA-Op) [device ID, activity schedule table]

Figure 16 shows CM functions and the interfaces.

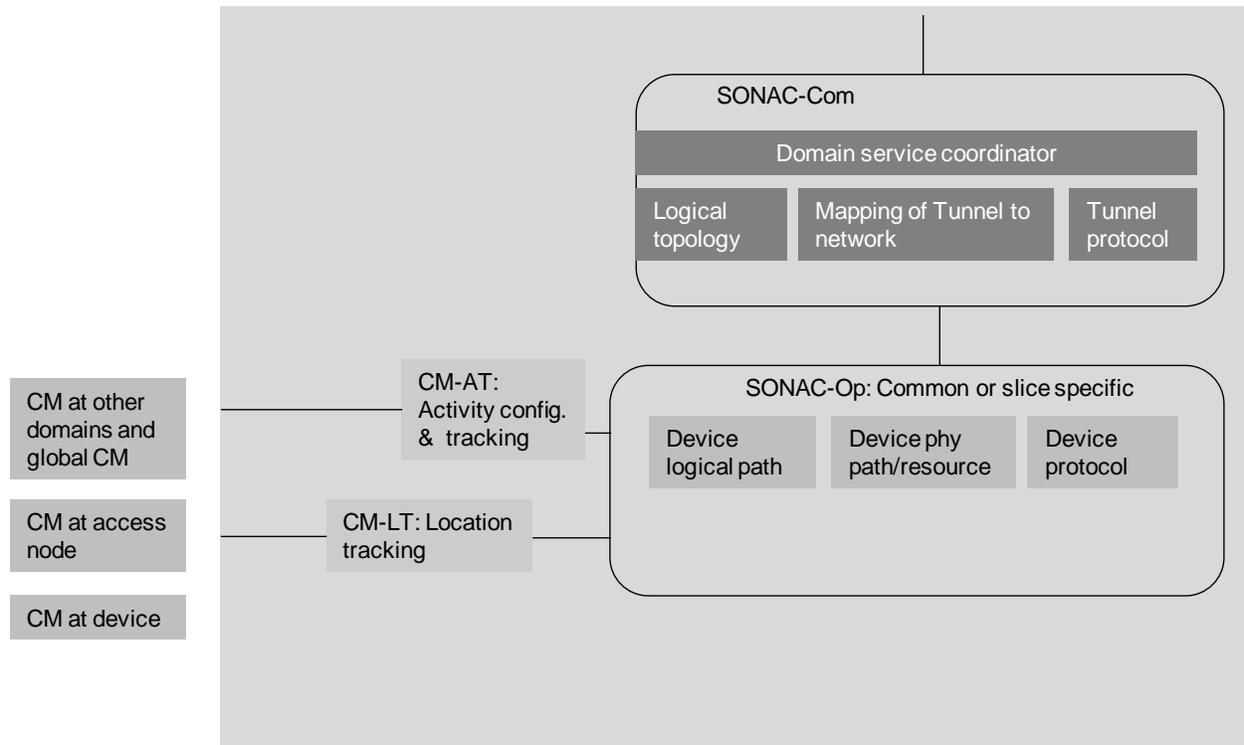

Figure 16. CM functions and interfaces.

## 6.7. CFM functions

In 3G/4G, the CDN (content distribution network) service is a type of OTT service which distributes the contents produced by content providers over the network. In the future, in addition to the CDN providers, content producers of various types may wish to share their contents publically or with a group of remote consumers.  Also when content is downloaded to a subscriber for the first time, the content can be cached by the network without violating agreements between the operator and the content producer. The future network, thus, can be viewed as a big content library. A slice provider can define a content cache and forwarding slice (content library slice) to enable the efficient content sharing. The content creation could be very dynamic and the way to forward these contents could be significantly different from that for CDN. CFM functions are defined for controlling and managing the content cache and forwarding (CF) slice. A CFM slice can also be created by a CF customer. Such a slice includes the caches distributed in the infrastructure network. The selected contents are kept in the selected caches.

*Functions:*

- Content registration management
- Configure DAM to monitor the probability of content requests

- Determines and manage content cache
- Register to upper layer CFM function and neighbor CFM functions when any new content is cached or any existing content be removed
- Maintain content cache MAP
- Indicate to SONAC-Op whenever receiving a content registration from other CFMs
- etc

*Interfaces:*

- DAM for content request log and analytics [reporting mode, log mode]
- Caches of the CF slice for cache instruction [content ID, cache or remove, etc]
- SONAC-Op (SDT-Op) [content ID, cache ID]
- Other CMF [content ID, VN Node ID (Cache ID), cached or removed]
- 3rd party [content cached] or [content cache MAP]
- etc

Figure 17 shows CFM functions and the interfaces.

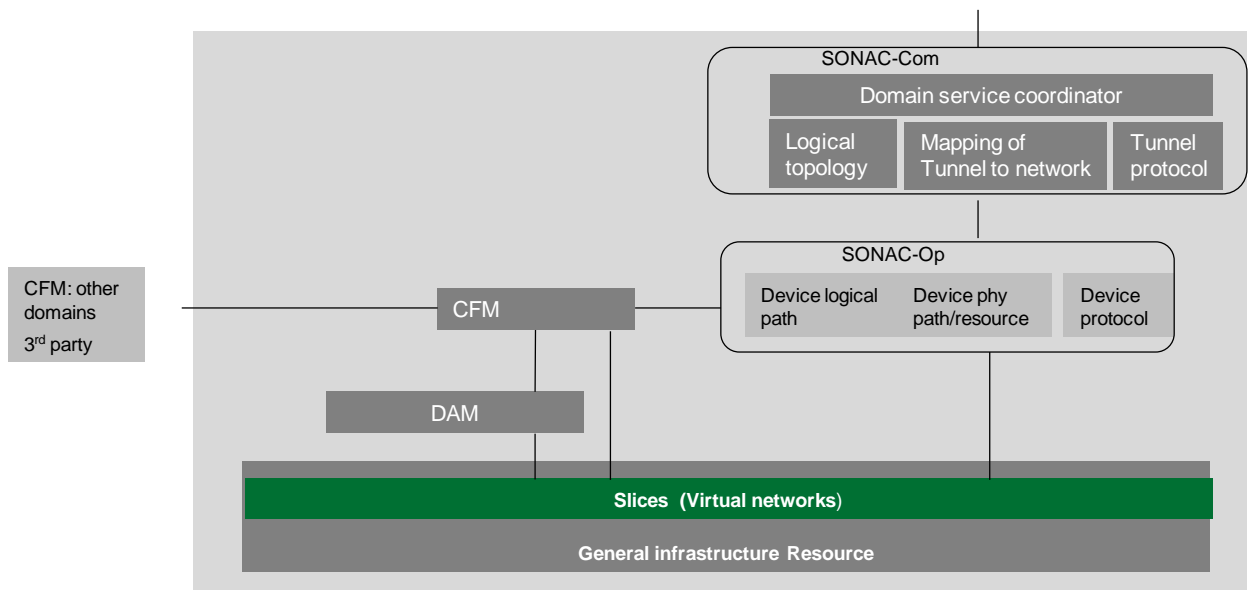

**Figure 17. CFM functions and interfaces.**

# 7. MyNET platform and the development of customer service slice

After MyNET platform has been developed and deployed, MyNET platform is ready to develop customer service slices.

A slice provider can also pre-deploy one or more 'purpose-specific slice'. Each of them is customized to a type of representative services, prior to receiving service requests from customers.

Depending on the resource configured, multiple formats of 'purpose specific slices' and customer service slices can be developed and deployed as listed in Table 1.

Table 1. SONC-Com and formats of slice resource assignment.

| Format ID | Slice topology and resource | Slice resource description | SONAC-Com responsibility - Creation of slices (purpose-specific slices or customer service slices) from infrastructure network | SONAC-Com responsibility - Creation of customer service slice from pre-deployed purpose-specific slice |
|---|---|---|---|---|
| A | VN node (NF) of a slice; Edge NN (e.g., access node); Device/server of a slice | • NFs definition/selection<br>• NFs deployment in DC/MEC<br>• Interconnections among NFs<br>• Capacity/performance requirement of logical tunnels/DCs/MEC<br>• Mapping logical tunnels to infrastructure network (mapping rule)<br>  - logical: IP-like, source routing, MPLS<br>  - dedicated physical resource<br>• End-to-end/tunnel/link protocols | • slice admission control described for format A slice<br>• define the slice resource<br>• deploy (embed) a slice into infrastructure network<br>• define/activate/configure<br>  - SONAC-Op – UP slice resource/per end-point QoS<br>  - NOS | • customer slice admission control<br>• optional resource assignment (format A) for the requesting slice without violating the capacity definitions of the purpose-specific slice |
| B | VN node (NF) of a slice; Edge NN (e.g., access node); Device/server of a slice | • NFs definition/selection<br>• NFs deployment in DC/MEC<br>• Interconnections among NFs<br>(no capacity/performance requirement of a slice)<br>• default mapping rule and default protocols<br>(assuming pre-defined default mapping rule and default protocols for slices) | • define the slice resource described for format B slice<br>  - NFs definition/selection<br>  - NFs deployment in DC/MEC<br>  - Interconnections among NFs<br>• deploy (embed) the slice into infrastructure network<br>• define/activate/configure<br>  - SONAC-Op<br>  - NOS | • Customer slice admission control<br>• optional resource assignment (format A) for a requesting slice |
| C | VN node (NF) of a slice; Edge NN (e.g., access node); Device/server of a slice | • NFs definition/selection<br>• NFs deployment in DC/MEC<br>(no topology definition of a slice) | • define the slice resource described for format C slice<br>  - NFs definition/selection<br>  - NFs deployment in DC/MEC<br>• deploy (embed) the slice into infrastructure network<br>• define/activate/configure<br>  - SONAC-Op<br>  - NOS | • customer slice admission control<br>• optional resource assignment (format A, B) for a requesting slice |

Format A slice: an end-to-end slice with the definition of NFs, slice logical topology, capacity and performance requirement of all logical tunnels of this slice, the mapping rule between logical topology to infrastructure network resource, the definition of end-to-end/tunnel/link protocols. There could be multiple mapping schemes between a logical tunnel and the infrastructure network for the purpose of traffic packets forwarding. For example, for soft mapping, e.g., IP- like routing, source routing, and destination based routing are possible; for hard mapping, e.g., dedicated resource assignment to a tunnels of a slice.

Format B slice: an end-to-end slice with the definition of NFs and the slice logical topology, but without the definition of capacity and performance requirement of all logical tunnels of this slice.

Format C slice: slice with only the definition of NFs and placements of NFs in the infrastructure network, but without the logical topology defined.

Variations of these formats are also possible.

A customer service slice can be created from a 'purpose-specific slice', and can also present different format.

The design of SONAC-Com enables the development of the follows:

- 'Purpose-specific slices' from the general infrastructure network
- Customer service slice directly from the general infrastructure network
- Customer service slice from pre-deployed 'purpose-specific slice'

## 7.1. Customer service slice development

The customer slice development procedure is highlighted in Figure 18. In this paper, a customer slice is a UP slice.

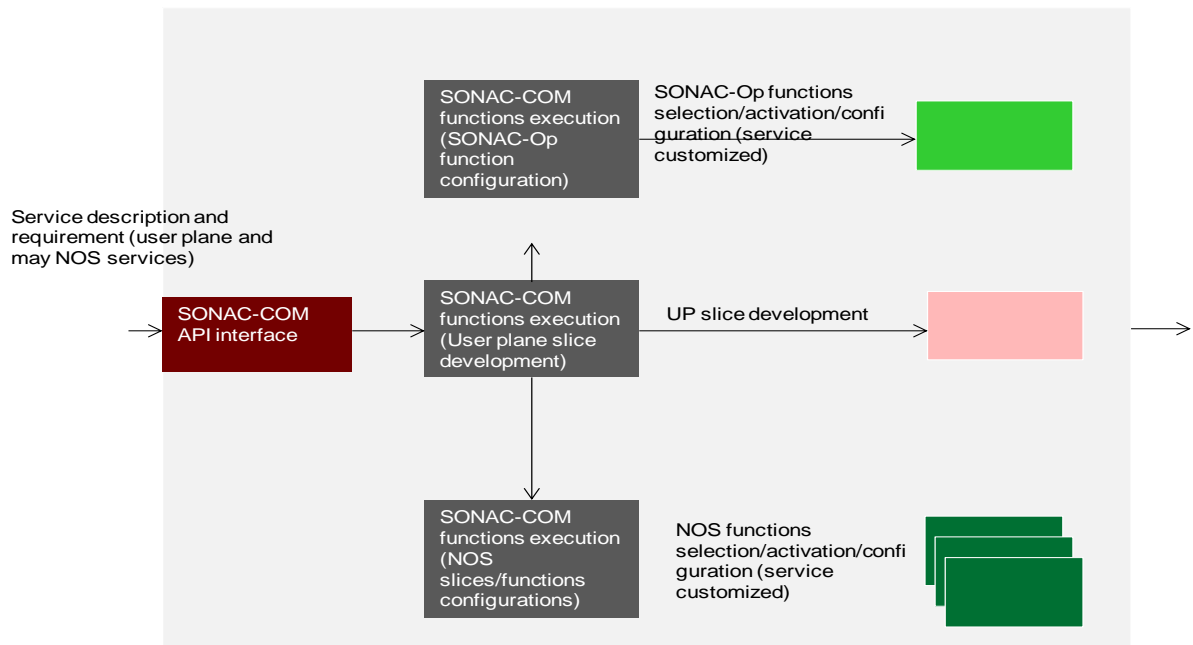

**Figure 18. SONAC-Com and customer service slice development.**

SONAC-Com receives the description and requirement from a slice requester (customer) via API interface. This information will be used by SONAC-Com to develop the UP slice by the execution of SONAC-Com functions. This procedure can also be used for slice providers to develop 'purpose-specific slices'. After the design of a slice, the design blueprint is realized (deployed) in the infrastructure network. ETSI NFV can be used for the instantiation of virtual functions in the selected cloud (e.g., DCs, MECs). The responsibility of SONAC-Com is summarized in Table 1.

After a UP slice is developed, SONAC-Com needs to select and/or activate and/or configure SONAC-Op functions by the execution of SONAC-Com functions for this purpose. If shared SONAC-Op functions are used for this UP slice, SONAC-COM only needs to configure the existing SONAC-Op functions for supporting the operation of the newly developed slice. Or SONAC-Com can select the service specific SONAC-Op functions, activate them and set corresponding operation rules/parameters of these functions.

NOS functions, e.g., CSM function needs to be configured to enable the management of the newly created slice. Similarly, CM functions need to be configured to prepare to provide reach-ability information of devices.

The inter-connections between SONAC-Op slice and NOS slices need to be determined.

After the development, deployment and configuration procedures, the customer devices can be admitted to these slices and service traffic data can be forwarded over UP slice and NOS signaling messages can be forwarded over NOS slices.

## 7.2. Multiple customer service sharing one UP slice

As discussed previously, a purpose-specific slice can be either dedicated to a single service customer or be shared among multiple customers if these services share the common service attributes and requirements. For example, all of smart reading services can share a UP slice ('smart reading slice'), data traffic of these services can be treated in the same way (QoS) by the network.

Considering this scenario, when a service request is received, SONAC-Com determines which domains (segments of a purpose-specific UP slice) will involve in the support of this service, and configures the corresponding SONAC-Op and NOS functions, as summarized in Table 1. SONAC-Com also needs to perform service admission control to make sure that the estimated resource requirement can be satisfied without causing any negative performance impact on other existing slices.

## 7.3. Development of slice with customized SONAC-Op and/or NOS functions

In some cases, a customer of a slice wishes to provide service to its subscribers. Slice owner may also explicitly require the customized SONAC-Op functions and/or NOS functions. The interface API can be used for this purpose. For this case, SONAC-Com develops a UP slices, a customized SONAC-OP slice, and/or NOS slices based on the description on the requirement of the slice operation.

## 7.4. Development of self-operational virtual network

Some slice customers require a set of infrastructure resource, including UP plane and functions in SONAC-Op and NOS functions, e.g., CM and CSM, and also want to operate their UP slices at the "virtual network level", i.e., a the level of the abstracted virtual network. SONAC-Com enables the development of such self-operational virtual network/slice. In this case the virtual network operator can access only the virtual network resource, without accessing the physical infrastructure resource.

## 7.5. Development of virtual network with dedicated end-to-end infrastructure resource

In some cases, a customer may request a set of dedicated infrastructure network resources in order to provide the communication service to its subscribers.

In this case, SONAC-Com simply allocates the dedicated end-to-end infrastructure network resources (hard slicing) based on the requirement of customers to create a non-sharable virtual network. Within the virtual network, all MyNET platform functions can be deployed. The dedicated infrastructure network resources can be viewed as the entitled infrastructure resource. In some cases, an end-to-end slice consists of both "hard" slice segments and "soft" slice segments. For example, a part of spectrum resource in access link segment is dedicated to a slice but no dedicated resources are allocated in other parts of the slice.

## 8. MyNET platform and the adaptation of customer service slices

One of SONAC-Com responsibilities is to enable slices to adapt to the real-time changes of certain conditions. This is the most challenging part of SONAC-Com design.

### 8.1. Factors impacting on the adaptation of a slice

A number of factors can trigger the slice adaptation, as shown in Figure 19.

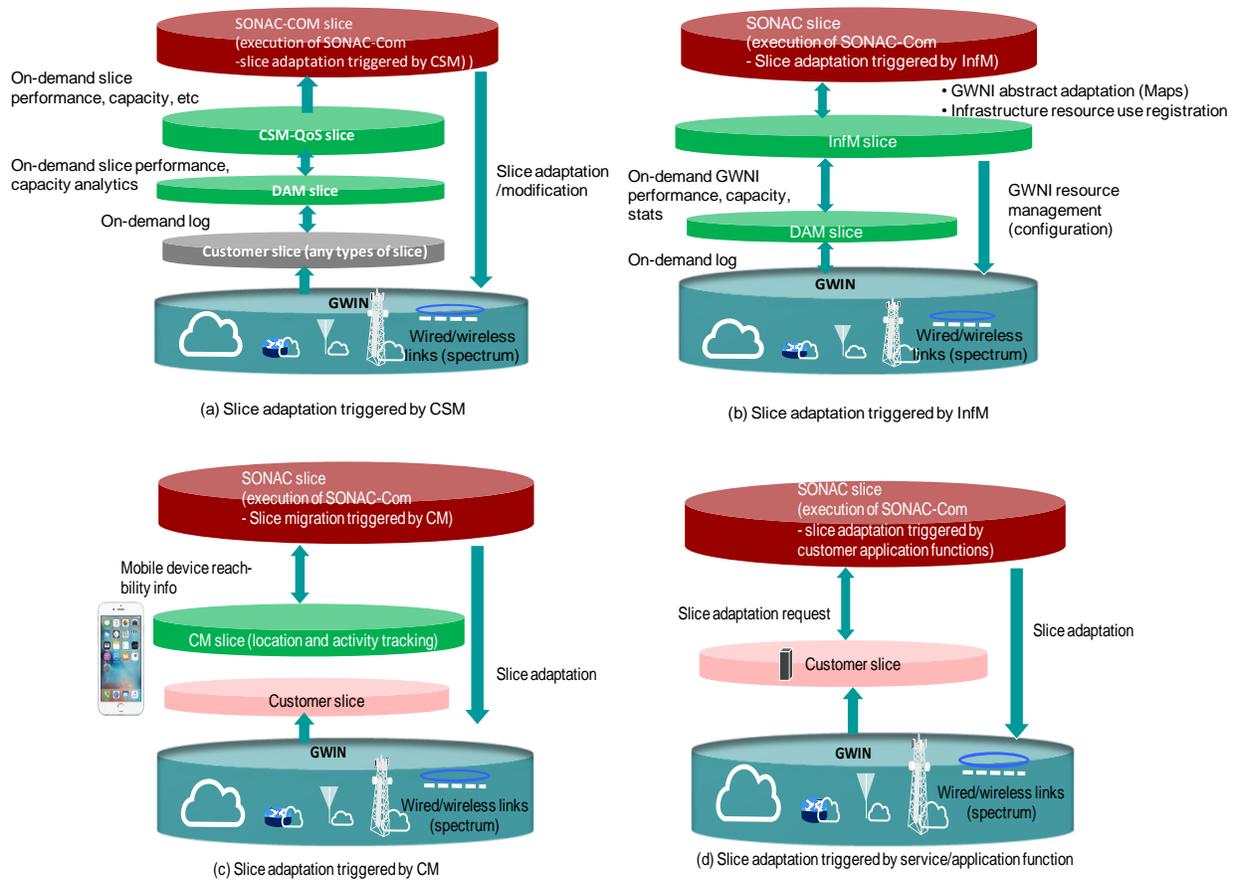

Figure 19. Factors impacting on the slice adaptation.

- Customer service assurance aspect – CSM-QoS

CSM-QoS is designed for the optimization of slice resource utilization and the assurance of slice service performance, as shown in Figure 19 (a).

From the customer service perspective, there are multiple reasons behind the adaptation of a slice:

- Although SONAC-Com algorithm is developed to try to best match the assigned resource to the needs, certain level mis-match can be expected. Whenever such mis-match is identified by CSM-QoS, the slice resource modification should be performed. The mis-match includes the over-provisioning and under-provisioning of the slice resources. For the over-provisioning of the slice resource, the infrastructure network resource may not be efficiently used. The under-provisioning of the slice resource may cause the unsatisfied slice service performance, i.e., the delivered service performance does not match that was promised by slice provider. These can be identified by CSM-QoS through the slice resource utilization statistics and slice performance statistics.
- The description of service attributes by slice requester may not be perfectly reflects the real service attributes. CSM-QoS needs to monitor the statistics of the pattern of traffic load, etc and correct the service description.
- For some service, by nature, it is difficult to provide accurate description due to the dynamics of the traffic load and device mobility, etc. This means that the slice resource allocation needs to be adapted to such dynamics after the initial deployment. CSM-QoS needs to continuously monitor the slice traffic load change and migration such that the slice resource allocation statistically matches the slice traffic load behavior. CSM-QoS triggers the slice adaptation if certain pre-configured conditions are met.

The adaptation of a slice to enable the slice service assurance and the optimization of the slice resource allocation is a close-loop scheme where CSM-QoS plays an important role.

- Infrastructure resource optimization aspect

The assurance of the infrastructure network performance and the optimization of the infrastructure network resources utilization can result in the changes in the Infrastructure network topology, capacity, etc. This can cause the adaptation of existing slices and be triggered by InfM functions, as shown in Figure 19 (b). There are a number of reasons that cause the changes.

- The performance degradation of the infrastructure network equipments/elements, i.e., equipment failure, etc. InfM is responsible for the performance assurance of infrastructure network by monitoring the infrastructure equipments and make decision on infrastructure network adaptation.
- The on-demand changes in the wireless backhaul network topology. One of the responsibilities of InfM is to determine the RAN backhaul network topology based on traffic load statistics, i.e., power-on/off some of wireless network nodes.
- The on-demand integration of the additional infrastructure networks. When the infrastructure network resources are not sufficient to admit new slice requests, InfM can trigger the integration and enlarge the size of the infrastructure network resource pool.

The changes in the infrastructure network topology, resource capacity, etc, result in the adaptations of some existing slices.

- Device service aspect

For some extreme cases where a customer slice is dedicated to a customer with mobile devices, e.g., an emergency ambulance service customer, where some service specific functions need to be deployed close to the edge of the network and need to migrate along with the mobile devices. In this case, the slice topology may be adapted to the movement of the mobile devices, i.e., the locations of the devices may trigger the adaptation of current slice. This is referred as the slice migration. CM triggers such a slice migration, as shown in Figure 19 (c).

- Customer service aspect

For some services, the traffic load distribution may adapt to certain event. This may require the slice adaptation to the event. Such an event can be detected by the service application function. In this case, the application function may trigger a slice adaptation, as shown in Figure 19 (d). The interface between such an application function and SONAC-Com may need to be configured at the initial slice deployment.

## 8.2. Slice adaptation

### 8.2.1. CSM-QoS for slice performance assurance

CSM-QoS functions are designed to make sure that the slice service performance is satisfied during the entire slice lifecycle, and ensure that the slice resource matches the slice traffic all the time.

As shown in Figure 19 (a), this is a close-loop controlling system.

After a slice is developed and deployed, CSM-QoS needs to be configured. The information of this newly created slice needs to be informed to CSM-QoS. This information includes the slice service description and service requirement. This information is used by CSM-QoS as certain thresholds to compare against in order to trigger a slice adaptation. This function is owned by slice providers.

For triggering a slice adaptation, CSM-QoS needs to send message (e.g., slice adaptation request) to SONAC-Com function which is responsible for the slice adaptation triggered by CSM, as described in Chapter 3. The message should include the following contents: slice ID, cause ID (e.g,, slice resource utilization statistics, slice service performance statistics, slice traffic behavior statistics), and the corresponding values.

After receiving the request, SONAC-Com determines whether the topology of the slice needs to be changed, or only slice resource modification (increase or reduce slice resource) are needed. After the adaptation of the customer service UP slice, SONAC-Com reconfigures NOS and SONAC-Op in case needed.

CSM-QoS function obtains the statistics of slice resource utilization, the slice service performance, and the long term slice traffic load behavior from DAM.

For the cases where over-provisioning of slice resource is identified, CSM-QoS may, as configured by SONAC-Com, inform the domain or end-to-end resource management function of SONAC-Com. This triggers SONAC-Com to modify the slice resource allocation.

For the cases where the mis-matching between the delivered slice performance and the promised performance is identified, CSM-QoS interacts with the domain service coordinator or the end-to-end service coordinator. This may trigger the modification of slice resource allocation.

For the cases where the long term statistics of traffic load distribution is different from what is provided by the customer, the statistic information may need to be sent to the customer and the service coordinator. This may results in the adjustment of the slice resource allocation.

For the cases where the traffic load, by nature, is difficult to estimate, CSM-QoS needs more frequently monitoring of the traffic load change. CSM-QoS reports the changes to SONAC-Com for performing a slice adaptation.

The operation of CSM-QoS is a close-loop controlling procedure. Slice performance assurance is one of key techniques in future wireless networks due to the introduction of slicing concepts.

### 8.2.2. InfM enabling infrastructure performance assurance

InfM function is designed for the optimization of infrastructure resource utilization. For InfM triggered slice adaptations, InfM sends message to SONAC-Com function which is responsible to process the request message received from InfM. The information carried in this message includes the information of infrastructure network updates.

SONAC-Com, after receiving this message, needs to determine the impacted slices, determine the adaptation of current slices, and reconfigure NOS and SONAC-Op if needed.

For the cases where a performance degradation of the infrastructure equipment/element is identified, InfM needs to take certain reaction, e.g., to activate a backup system. At the same time, if this even causes a change in the infrastructure topology, capacity, etc, InfM needs to inform the domain SONAC-Com and the end-to-end SONAC-Com by providing the updated infrastructure network information.

For the cases where the infrastructure resource utilization rate is low, InfM should react to this by de-activate certain infrastructure elements, or some sub-systems of an equipment. Such a change needs to be informed to SONAC-Com.

For the cases where the rejection rate of slice requests is higher than a pre-configured threshold, InfM may trigger the integration of additional infrastructure networks. This updates needs to be informed to SONAC-Com.

For the cases where the access link frame structure needs to be modified based on traffic load statistics, InfM needs to inform SOMAC-Com and SONAC-Op.

It should be noted that in order to make smooth adaptations of the existing slices due to the changes of the infrastructure networks, InfM and SONAC-Com need to collaborate closely to avoid any interruptions of the operation of the existing slices.

### 8.2.3. Service application triggered slice adaptation

For services which require the slice adaptation triggered by certain event during the slice operation, application layer functions are used to identify such a event and is allow to interact with SONAC-Com to trigger a slice adaptation. The interface between the application function and SONAC-Com should be defined (or reuse the API interface at the slice initialization). The message requesting a slice adaptation may carry information such as, slice ID/service ID, the service descriptions, the service requirement, etc.

SONAC-Com function which is responsible for the process of the request message from the service application function needs to determine the impacted segment of the slice, determine the adaptation of the current slice and reconfigure NOS and SONAC-Op if needed.

### 8.2.4. CM triggered slice adaptation

For certain cases where a slice is dedicatedly designed for a customer with certain number of mobile devices and with service specific functions implemented in the slice, the migration of devices may cause the slice migration, i.e., application function is migrated from one place to another place. Such a migration may be triggered by CM function which is responsible for performing the device location tracking.

To enable this, CM function needs to be configured to trigger slice migration if certain condition is met. The slice migration request message sent by CM function to SONAC-Com should include information, such as, slice ID, mobile device migration information, etc.

SONAC-Com function which is responsible for processing slice migration request needs to determine the adaptation of the current slice. This may include new function instantiation or activation or configuration of pre-installed functions, reconfiguration of NOS and SONAC-Op if needed.

### 8.2.5. CFM for content delivery optimization

For a CF (Content and Forwarding) slice, based on the statistics of the requests of certain contents, CFM may trigger an existing CF slice to adapt to the detected change, e.g., to instantiate caches close to edge of the networks to reduce downloading latency of certain contents. CFM function needs to send a request message to SONAC-Com for this purpose. This message may carry information, such as, device distributions which are expected to be interested in certain contents, and the performance requirement of content delivery.

SONAC-Com function which is responsible for processing such a request from CFM function needs to determine the cache location and adapts the slice topology.

## 8.3. MyNET platform and slice adaptation

Figure 20 shows MyNET platform and the interactions among MyNET functions to enable slice adaptation. SOMAC-Com needs to interface with CSM-QoS, InfM, CM, CFM and customer service application functions.

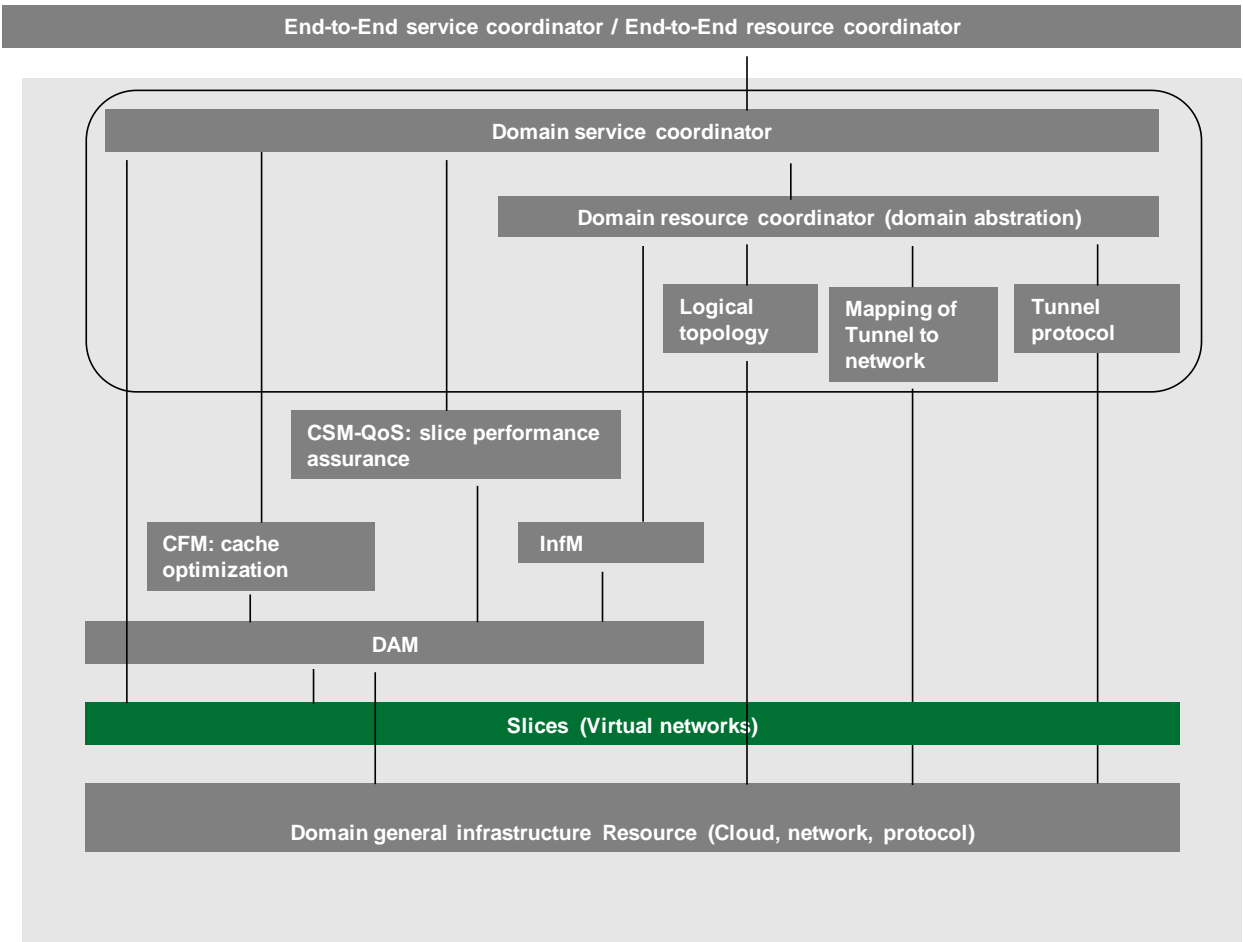

Figure 20. MyNET platform and slice adaptation.

## 8.4. Scope of adaptation

The adaptation of a slice may be end-to-end or only local adaptation. For local adaptation, i.e., intra-domain adaptation, the adaptation can be invisible to other domains if there is no impact on other domains. A domain SONAC-Com should make a decision on whether this adaptation impacts on other domains or not after analyzing the information provided by InfM or CSM or application functions in the same domain. If the adaptation impacts on more than one domain, the end-to-end SONAC-Com may involve in the adaptation.

# 9. MyNET platform and the operation of slices

SONAC-Com and SONAC-Op are designed for the automation of slice creation and slice operation. The resource defined for a slice by SOMAC-Com could present multiple formats, as discussed in Chapter 7. SONAC-Op is responsible for the slice operation. The responsibilities of SONAC-Op for different formats of slice design are different as described in Table 2.

**Table 2. SONAC-Op and formats of slice resource assignment.**

| Format ID | Slice topology and resource | SONAC-Com responsibility | SONAC-Op responsibility | Note |
|---|---|---|---|---|
| A | VN node (NF) of a slice; Edge NN (e.g., access node); Device/server of a slice | • define the slice resource described for format A slice<br>• deploy (embed) the slice into infrastructure network<br>• define/activate/configure<br>  - SONAC-Op<br>  - NOS | • customer/device network and slice registration<br>• device traffic session admission control<br>• **routing data packet over slice**<br>• **interact with CM for mobility**<br>• device data packet path selection on a logical tunnel if needed<br>• link (e.g., access link) resource multiplexing control without violating slice performance requirement<br>• device data packet protocol determination without violating the per slice protocol selection principle<br>• only access link connection resource control signaling | • more intelligent SONAC-Com<br>• simple SONAC-Op<br>• easy slice QoS control<br>• easy global congestion control<br>• allow customized mapping rule and protocol<br>• no need for end-to-end PDU session establishment for devices<br>(save signaling during the operation phase) |
| B | VN node (NF) of a slice; Edge NN (e.g., access node); Device/server of a slice | • define the slice resource described for format B slice<br>• deploy (embed) the slice into infrastructure network<br>• define/activate/configure<br>  - SONAC-Op<br>  - NOS | • customer/device network and slice registration<br>• device traffic session admission control<br>• **routing data packet over slice**<br>• **interact with CM for mobility**<br>• device data packet path selection on a logical tunnel if needed<br>• device data packet protocol determination without violating the per slice protocol selection principle<br>• access link connection resource control signaling | • Less intelligent SONAC-Com<br>• More intelligent SONAC-Op<br>• less customization of slice<br>• more complicated slice QoS control<br>• no need for end-to-end PDU session establishment for devices |
| C | VN node (NF) of a slice; Edge NN (e.g., access node); Device/server of a slice | • define the slice resource described for format C slice<br>• deploy (embed) the slice into infrastructure network<br>• define/activate/configure<br>  - SONAC-Op<br>  - NOS | • customer/device network and slice registration<br>• device traffic session admission control<br>• **per device end-to-end session establishment**<br>• **interaction with CM for mobility – end-to-end session re-establishment**<br>• device data packet path selection on a logical tunnel if needed<br>• device data packet protocol determination without violating the per slice protocol selection principle<br>• end-to-end device PDU session establishment and re-establishment signaling<br>• access link connection resource control signaling | • Less intelligent SONAC-Com<br>• More intelligent SONAC-Op<br>• need per device end-to-end (incur more real-time signaling during slice operation) |

## 9.1. Responsibility of SONAC-Op for the operation of format A slice

After a slice is completely configured and developed, the slice can be used or operated for forwarding traffic of end-points (devices, servers, etc) of the slice. SONAC-Op could be common for more than one slice or dedicated for one particular slice, e.g., cache and forwarding (CF) slice. At the development of a customer service slice, SONAC-Com also configures SONAC-Op functions to have them to control and manage the traffic handling of a slice. With the deployment of an end-to-end slice, all involved network nodes should 'know' how to handle data packets of this slice. The operation of such a slice should be simplified. For example, per device end-to-end session establishment is not needed due to the well defined slice. Only the connection between a device and access node(s) is needed which needs much less signaling. Therefore the singling overhead and the latency due to the per device session establishment can be reduced during the slice operation. The mobile device data transmission to the network becomes to 'select' a right 'enter point' of a slice during the movement. The DL data delivery to a mobile device from the network becomes to 'select' a right 'exit point' of a slice. An end-point of a slice simple transmits data on to the slice with no or only lightweight AL signaling for access link resource acquisition. This feature is called as Hop-On (a Slice) and will reduce the complexity (signaling overhead and latency) of data traffic handling during the operation of a slice [11].

In order to use a slice, an end-point of a slice (device, servers) needs to register to the slice. After the registration, an end-point device should obtain and maintain an identification (slice ID, service ID, etc) for the network to associate packets to a right slice.

When a device sends packets to another device, the sender simply inserts the name of the targeted device without worrying the physical location of the destination device. After a packet is received, SDT-Op, as a virtual router, needs to route the packet to the destination over a slice. For routing packets over a slice (slice tunnels), SDT-Op needs to have the location information of the destination of the packets. This information will be obtained from CM based on the name of the destination device. After SDT-Op determines the forwarding tunnel, SDRA-Op at the ingress of the tunnel will instruct the associated physical node to physically move packets from the ingress to the egress of the logical tunnel. The access link scheduler (SDRA-Op at access node) needs to perform the access link resource assignment for devices based on the policy configured by SOMAC-Com at the development of the UP slice. SDP-Op manages the protocol stack selection for devices if the default stack of a slice is not optimized.

During the slice operation, SDT-Op needs to interact with CM to maintain the end-points (devices, servers) routing table. SDRA-op needs to interact with CSM-context for the resource multiplexing of end-points to optimize the slice resource usage. CSM-AAA Controls and manages the authentication and authorization of devices to use the slice.

Due to the hierarchical architecture of both SONAC-Op and CM service, the local routing on a slice can be easily supported. For example, if a device sends packets to another device in the same domain as the sender device, the packets can be routed to the destination devices without going out the domain. The local routing feature is enabled by SONAC-Op and CM.

The strategy of designing a format A slice is to move the complexity during the real-time operation of a slice to the phase of the slice development and deployment. The operation of a format A slice can be viewed as a combination of the circuit oriented switch techniques and the packet-oriented routing techniques. The slice design defines a slice logical topology (virtual network) and SDT-Op work just like an IP routers but working on the logical topology level with much simple local intelligence.

## 9.2. Responsibility of SONAC-Op for operation of format B slice

The responsibility of SONAC-Op for format B slice is similar to that for format A slice. However, since there is no slice level capacity is defined, the slice level QoS control may be difficult. For example, scheduler at access link has the difficulty to ensure the resource multiplexing 'fairness' among multiple slices.

## 9.3. Responsibility of SONAC-Op for operation of format C slice

For format C slice, during the operation, the responsibility of SDT-Op becomes to perform per device end-to-end session establishment. This incurs signaling between SDT-Op and UP plane, i.e., involved UP NFs during the operation of a slice. For a slice associated with a high number of devices and handling high bursty traffic, the significantly high signaling overhead could be expected. More intelligence in the SDT-Op function is needed. For a format C slice, a relative simple SONCA-Com function can be expected for the development of the slice. However, the more intelligent SONAC-Op functions are needed and more signaling overhead and latency can be predicted.

For different formats of slices, the operation of a slice could be different. This requires an explicit description of the slice resource assignment and corresponding configuration of SONAC-Op. The format of a slice is determined by slice providers based on the considerations, such as the purpose of a slice, the complexity balance between SONAC-Com and SONAC-Op, etc.

The design of MyNET architecture enables the flexible balance between the complexity of the slice definition/design and the complexity of the operation of a slice.

## 10. MyNET platform and the business revenue assurance

In 3G/4G, the business model is relative simple. The network operator provides the communication services to their subscribers. The operators log per subscriber's traffic for the purpose of charging. The log function is located in the PGWs and the charging granularity could be at per device level or per application level. Both on-line and offline charging are defined.

In future networks, multiple types of players can be expected as discussed in Chapter 1. Some of examples of future network service provisioning models are shown in Figure 21 (a) and Figure 21 (b). These figures only capture certain models and a lot of different models can be expected.

For some cases, one infrastructure provider can provide its infrastructure network resource to multiple slice providers, as shown in Figure 21 (a). One slice provider can provide multiple slices using the infrastructure network resource. One slice can provide services to one or multiple customers. One slice customer can associate with one or multiple devices and servers which consume the slice resource for the communication purpose. For example, a smart reading customer can use a slice to collect reading messages from devices/sensors to a central server. Another type of slice customer could be a slice owner who can further provide service to its subscribers that actively consume the slice resource for the purpose of communications. For example, a virtual operator can use a slice to provide best effort service to its subscribers.

For some other cases, one slice provider can integrate infrastructure network resources from multiple infrastructure network providers, as shown in Figure 21 (b).

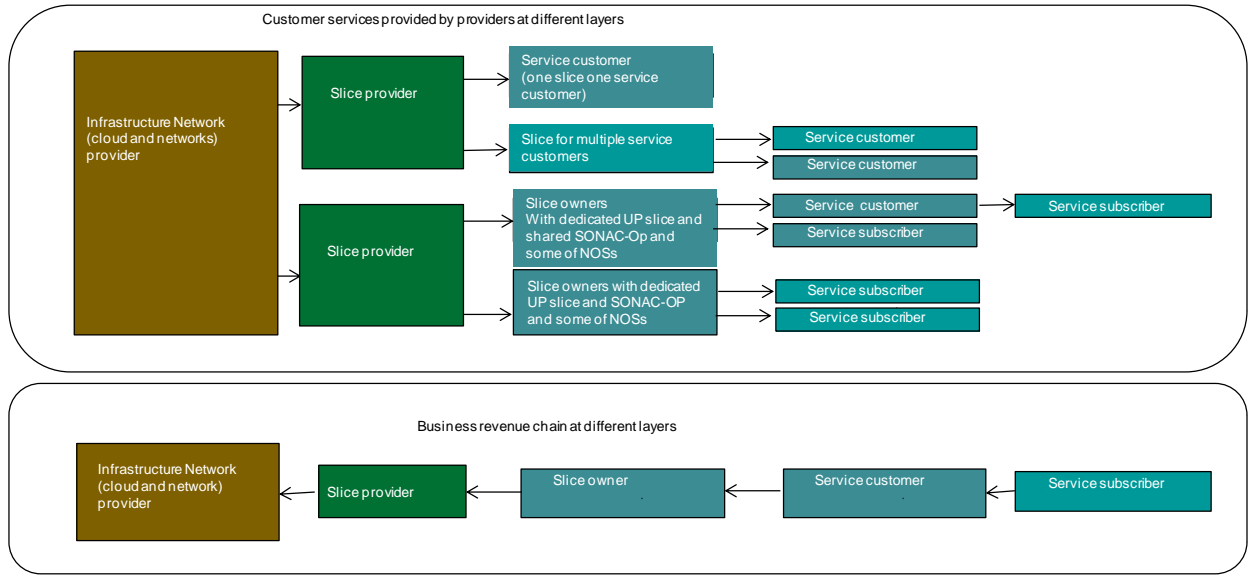

(a) Single infrastructure provider and multiple end-to-end slice providers.

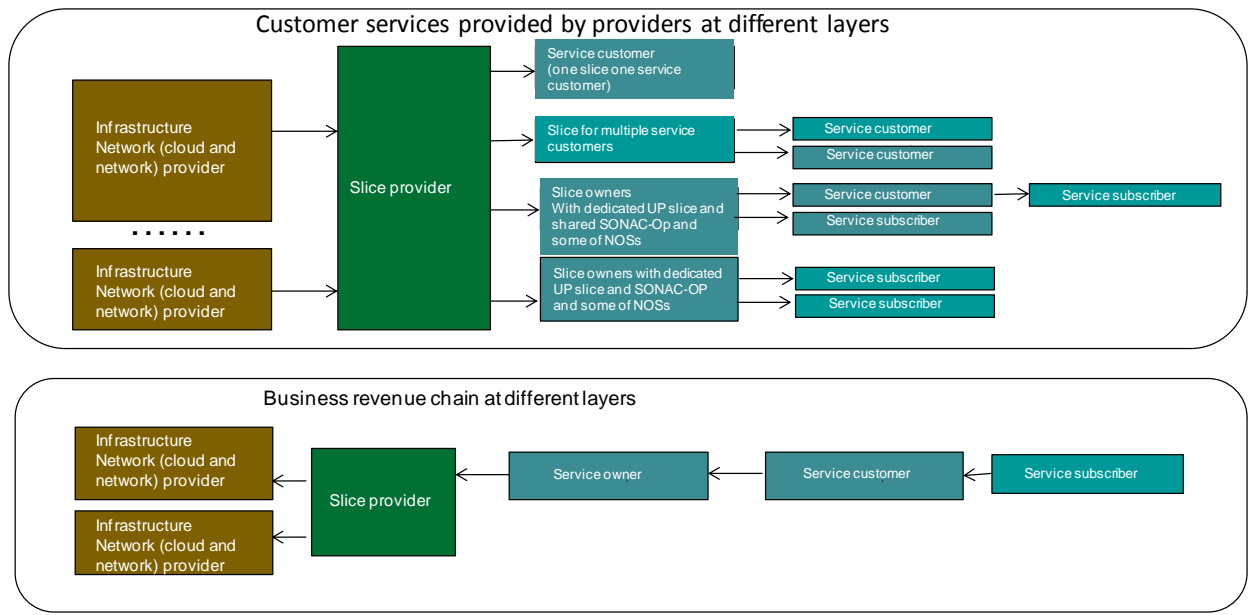

(b) Multiple infrastructure network providers and one end-to-end slice provider.

**Figure 21. Business revenue mode.**

From these example models, it can be predicted that the business chain models in future network industry could be significantly different from that in 3G/4G network. The new business revenue assurance scheme needs to be designed to meet the requirement of the new business chain model. The traditional charging scheme found in 3G/4G need to be enhanced to adapt to the new business models.

The following requirement needs to be considered for the business revenue assurance:

- Enable multiple level changing granularities
    - Enable charging at per application level
    - Enable charging at per subscriber level
    - Enable charging at per service customer level
    - Enable charging at per slice provider level
- Enable negotiable and service customized charging rule
- Enable flexible charging schemes
    - UP slice only charging
    - UP slice charging and SONAC-Op charging if it is dedicated to one customer
    - UP slice charging, SONAC-Op changing, and NOS slice charging (e.g., CM, CSM) if they are dedicated to one customer

The flexible and customized charging is enabled by MyNET platform – CSM-charging functions. CSM-charging function is configured when a customer slice is deployed. A customer can explicitly indicate the preferred charging rule and negotiate an agreed charging rule with the slice provider. During the operation, CSM-charging function configures DAM to obtain the information needed for the purpose of the assurance of business revenue of tall players.

After the development of a customer service slice, the information of the slice, such as, logical topology, needs to be provided to CSM-Charging function. This information can help CSM-Charging function in determining on where to log data for the purpose of charging.

The charging information obtained for each of the individual slices and services can be used for deriving the charging data at slice provider level.

# 11. MyNET platform and the Openness

In the future wireless network industry, highly efficient collaboration among multiple players at different layers is expected, as shown in Figure 22. One open network control and management platform is required. MyNET is designed to provide an open environment in order to enable benign collaboration among all players.

## 11.1. Slice/service customer aspect

From the customer perspective, a platform design needs consider the following aspects:

- Enable slice/service customers to be aware of the available infrastructure network information (topology, capacity, performance, etc) such that customers can actively be involved in the slice development under the control of slice provider.

- Enable slice/service customers to be aware of the real-time status of its slice/service (slice service performance, slice resource utilization, slice traffic load, etc).
- Enable slice/service customers/owners to be aware of the status of the behavior (applications, activity, commonly communicated partners, etc) of their devices or subscribers.
- Etc

In MyNET platform, InfM provides the status of the infrastructure network to SOMAC-Com for the purpose of the slice development. The status information can also be accessed by the slice customers, under the control of the slice providers and infrastructure providers.

The status of a slice or a service, provided by CSM, is used by SONAC-Com for the purpose of the optimization of the slice service performance and slice resource utilization. This status information can also be accessed by the customer of a slice, under the control of the slice providers.

The status of a device or a subscriber of a slice, provided by CSM and CM, is used for the optimization of resource control and management of slices. This status information can be accessed by the slice customers, under the control of slice providers.

3$^{rd}$ party can request DAM for certain information extraction under the control of the slice providers.

Figure 22 summarizes the support of the openness by MyNET platform.

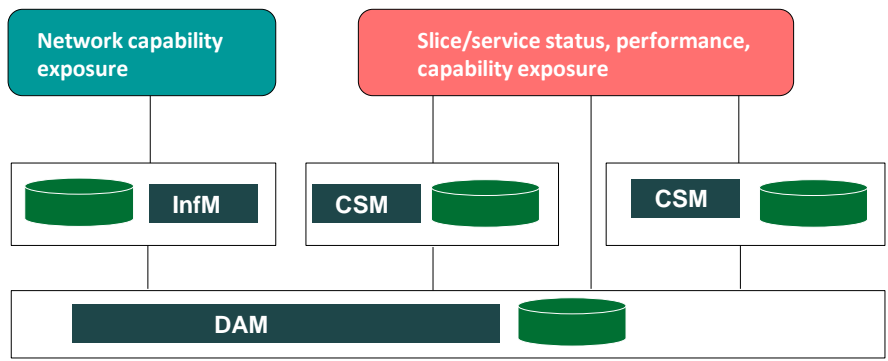

Figure 22. MyNET platform and the support of the openness.

## 11.2. Network control, management and operation aspect

From future network control, management and operation perspective, an architecture platform design should enable 3$^{rd}$ parties to control, manage and operate parts of a network under the control of the slice provider.

In the design of MyNet platform, the basic functions of the network control, management, and operation are indentified and categorized. The methodology of MyNET design makes MyNET to become a flexible open platform.

Due to the function classification of MyNET platform, each of the function families in MyNET platform is relatively independent from each other. This enables the openness of the network control, management and operation.

For example, DAM service could be provided by a 3$^{rd}$ party which is dedicated to the data log and information extraction.  DAM service provides the on-demand service to other MyNET functions. CM service can also be provided by a 3$^{rd}$ party which is dedicated to provide the service of the reach-ability of devices.

The openness can be at an end-to-end level or domain level. For example, for an end-to-end openness, one 3$^{rd}$ party DAM provider can provide the service over the entire infrastructure network; for a domain-wise openness, one 3$^{rd}$ party DAM provider may only provide DAM service over a limited part (e.g., a domain) of the entire infrastructure network.

## 12. Summary

In this paper, a future wireless network architecture, MyNET, is discussed.

SONAC is the key intelligent component of MyNET platform. SONAC-Com is responsible for the infrastructure resource multiplexing among multiple 'soft' slices and 'hard' slices. From the scope perspective, a slice means an end-to-end slice. The definition of a slice considers all the involved domains, including RAN, core, DCs and transport networks. From the resource allocation perspective, an end-to-end slice definition considers the slice topology, the mapping of the slice topology to the infrastructure network, and the data plane protocols. SONAC-Com is the 'brain' of MyNET platform and it should be designed to have the intelligence to adapt various 'conditions' predicted by human, and the capability to learn how to make 'decision' to deal with unpredicted 'conditions'. This capability needs to be programmed by men. This will become one of important research topics in the future.  SONAC-Op is responsible for the real-time traffic data packet handling over deployed slices. Given a well defined slice, the operation of a slice should be simplified. Slice providers can determine the format of a slice for different types of services and for different balances of the complexity between SONAC-Com and SONAC-Op.

The responsibility of InfM is to enable the optimization of the infrastructure resource pool utilization. InfM is designed to enable the integration and release of pieces of infrastructure resources, e.g., DCs, private networks, spectrum, mobiles/cars as temporary relay nodes. InfM enables an 'elastic' infrastructure network based on traffic load statistics, etc. For the infrastructure network which is composed of multiple domains, InfM needs to provide domain resource abstraction to SONAC-Com in some cases.

CSM is introduced for the management of a slice after the slice is deployed. This is a new function family. The key responsibilities of CSM are to ensure that a slice is really providing the required service quality; to ensure that the resource of a slice really matches to the traffic load.

CM is designed for providing the information of device reach-ability such that SONAC-Op can forward traffic packets to devices. The key concept is to decouple the name and the location of a device to avoid the change of the 'name' of a device due to the mobility. The multiple-level CM management also makes the reach-ability management scalable.

DAM (data log and analytics management) is introduced into the future network architecture to support the operation of SONAC and other NOS services. In addition, DAM can also provide service to other 3$^{rd}$ parties.

The responsibility of CFM is for the optimization of 'content' management, i.e., content registration, content cache and content sharing among parties who are interested to share their contents and who are interested to access the contents.

The key functions of MyNET platform are summarized in Table 3.

Table 3. Key functions of MyNET platform.

| Functions | MyNET PLatform |
|---|---|
| • end-to-end slice definition/realization (infrastructure resource multiplexing among slices)<br>   - end-to-end service coordination<br>   - end-to-end resource coordination/management | • SONAC-Com |
| • slice management (after a slice is realized)<br>   - slice/service/subscriber performance assurance<br>   - slice resource match slice traffic load | • SONAC-Com<br>• CSM |
| • slice operation (after a slice is realized)<br>   - traffic handling on a slice | • SONAC-Op<br>• CM/CSM |
| • infrastructure network resource management<br>   - infrastructure network performance assurance<br>   - infrastructure resource utilization optimization | • InfM |
| • device/mobile reach-ability management<br>   - device name and location decoupling<br>   - name based location tracking/resolution, activity status configuration and tracking<br>   - service customized reach-ability support | • CM |
| • data log and analytics (physical network elements and virtual function elements)<br>   - provisioning of DAM service to other functions | • DAM |
| • content (video, image, web page, etc) management<br>   - management of registration of contents provided by content service providers, end-customers, and any other entities<br>   - content cache optimization<br>   - content sharing/delivery optimization | • CFM |
| • Charging services<br>   - customized charging rule<br>   - multiple level charging – subscriber / service customer / slice customer / slice provider | • CSM |

It is expected that the future network architecture should be able to provide a number of capabilities, as discussed in Chapter 1. The capabilities of the future wireless network provided by MyNET are summarized in Table 4.

Table 4. Capabilities enabled by MyNET platform.

| Capability | MyNET Platform |
|---|---|
| Full automation of slice definition and realization | SONAC-Com |
| Full automation of performance assurance and resource elastics of slices | SONAC-Com/CSM |
| Full automation of infrastructure network resource elastics | InfM |
| Slice operation (traffic handling over slices)<br>- Hop-On - Not need for device end-to-end session establishment | SONAC-Com, SONAC-Op, CM |
| Slice operation (traffic handling over slices)<br>- Local routing | SONAC-OP, CM |
| Slice operation (traffic handling over slices)<br>- Roaming - free / simplified roaming | SONAC-Op, CM |
| Global congestion control | SONAC-Com, InfM, CSM-QoS |
| Name based reach-ability - location/activity and routing | CM |
| Service customized reach-ability | CM |
| Service customized security | CSM |
| Unified slice development of customer slices and wireless network operation services (NOS) | SONAC-Com, NOS |
| Unified management of contents provided by CDN providers and end-customers | CFM |
| Unified data log and analytics from both physical equipments and software functions | DAM |
| Business revenue assurance | CSM |
| Open environment of wireless operation and management | DAM, CSM, InfM |

Among those capabilities listed in Table 4, some of the capabilities are enabled by multiple MyNET components. For example, the global congestion control capability is enabled by SONAC-Com, CSM and InfM. In 3G/4G, the IP network uses the intelligent routing management protocol to enable the control of network congestion. The TCP protocol is used for end-to-end congestion control. These are 'distributed' schemes of the congestion control. In the future network, a more centralized and collaborated congestion control can be implemented. At the development of a slice, SONAC-Com function allocates slice resource based on the available infrastructure resource. This can avoid the traffic congestion in certain degree. During the slice operation, CSM continuously monitors the slice resource utilization and the slice service performance. Any slice performance degradation due to the slice traffic congestion should be reported to SONAC-Com. SONAC-Com then can modify the slice topology and capacity requirement of slice tunnels. Since SONAC-Com has a 'big picture' of the capacity of the infrastructure resource and the status of the available network resource, a better congestion control performance can be expected.

In this paper, MyNET platform is described in details. In the future, significant efforts will be put on the design of the identified intelligent functions in MyNET platform. This is the most challenging task and will involve the knowledge of communications and other theories as well, such as that of cybernetics.

With the development of modern sciences and technologies, and with the better understanding of the future wireless networks, designers of the future wireless network architecture will make the future wireless network a huge information exchanging and sharing "reservoir", and an indispensible part of everyday life. It will be future-proof as well. It is believed that such a network (architecture) will bring huge benefits to all of the players in the wireless network industry.